\documentclass[10pt,twocolumn,letterpaper]{article}
\usepackage{xcolor}
\usepackage[pagenumbers]{wacv} 

\definecolor{wacvblue}{rgb}{0.21,0.49,0.74}
\usepackage[pagebackref,breaklinks,colorlinks,allcolors=wacvblue]{hyperref}

\usepackage{tikz}
\usepackage{multirow}

\makeatother

\def\wacvPaperID{2540} 
\def\confName{WACV}
\def\confYear{2026}

\title{SuperRivolution: Fine-Scale Rivers from Coarse Temporal Satellite Imagery}

\author{
Rangel Daroya  \qquad
Subhransu Maji \\
University of Massachusetts, Amherst \\
{\tt\small \{rdaroya, smaji\}@umass.edu}
}

\begin{document}

\twocolumn[{%
\renewcommand\twocolumn[1][]{#1}%
\maketitle\begin{center}
    \centering
    \captionsetup{type=figure}
    \includegraphics[width=0.95\linewidth]{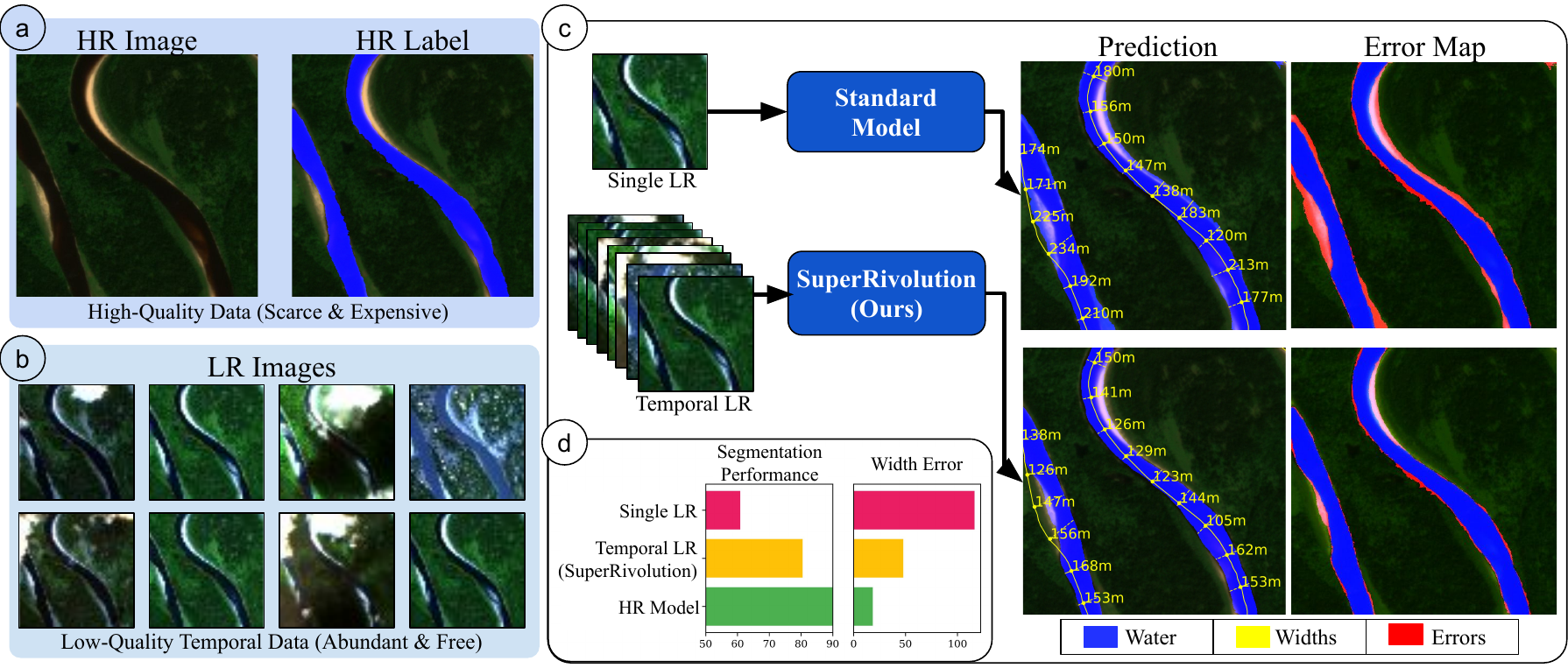}
    \vspace{-10pt}
    \captionof{figure}{\textbf{SuperRivolution leverages abundant low-resolution temporal data to improve performance on river segmentation.} (a) High-resolution (HR) data are precise but scarce and expensive to acquire. (b) Low-resolution (LR) satellite imagery are freely available and abundant, although of lower quality. (c) A standard model using a single LR image produces predictions with significant errors (top row, errors in red). (d) By fusing information from multiple LR images, our proposed approach generates significantly more accurate segmentations and more reliable river width estimates, reducing prediction errors and closing the gap with HR models.
    }
    \label{fig:motivation}
\end{center}%
}]

\begin{abstract}
Satellite missions provide valuable optical data for monitoring rivers at diverse spatial and temporal scales. However, accessibility remains a challenge: high-resolution imagery is ideal for fine-grained monitoring but is typically scarce and expensive compared to low-resolution imagery. To address this gap, we introduce SuperRivolution, a framework that improves river segmentation resolution by leveraging information from time series of low-resolution satellite images.
We contribute a new benchmark dataset of 9,810 low-resolution temporal images paired with high-resolution labels from an existing river monitoring dataset. Using this benchmark, we investigate multiple strategies for river segmentation, including ensembling single-image models, applying image super-resolution, and developing end-to-end models trained on temporal sequences.
SuperRivolution significantly outperforms single-image methods and baseline temporal approaches, narrowing the gap with supervised high-resolution models. For example, the F1 score for river segmentation improves from 60.9\% to 80.5\%, while the state-of-the-art model operating on high-resolution images achieves 94.1\%. Similar improvements are also observed in river width estimation tasks. Our results highlight the potential of publicly available low-resolution satellite archives for fine-scale river monitoring.
\end{abstract}

\vspace{-1em}
\section{Introduction}
\label{sec:intro}
Rivers play a central role in generating hydropower~\cite{kuriqi2021ecological,anderson2015impacts}, regulating water supplies~\cite{wurbs2005modeling,chen2005water}, and supporting diverse ecosystems~\cite{li2022integrating,patil2018understanding}. Monitoring these systems is therefore essential. Ground-based gauges provide direct measurements but are costly to install, require regular maintenance, and offer only sparse spatial coverage~\cite{ad2001global,gleason2017crossing}. In contrast, satellite remote sensing has emerged as a scalable alternative, delivering globally consistent optical observations that enable frequent and wide-area monitoring.

Spatiotemporal resolution is critical for river monitoring. Rivers are narrow structures with complex, dynamically evolving morphologies and surface textures, which makes their detection challenging~\cite{isikdogan2019seeing, flores2024mapping}. Many key hydrological parameters, such as discharge (\ie the volume of water flowing through a river), are not directly observable but must be inferred from physical principles using measurements of slope, width, and surface area~\cite{bjerklie2018satellite, hagemann2017bam}. Unlike conventional land-cover mapping tasks, hydrological monitoring therefore benefits substantially from high-resolution satellite imagery~\cite{daroya2025riverscope, tayer2023improving, kim2024evaluation}. 

High-resolution satellite imagery, however, is expensive to acquire. For instance, PlanetScope~\cite{planetlabs} provides commercial imagery at 3 m/pixel resolution, whereas publicly available missions such as Landsat~\cite{eros2020_landsat_l2_c2} and Sentinel~\cite{esa2022sentinel} offer lower-resolution data. Prior studies have shown that state-of-the-art models achieve a mean error of 116.4 meters on low-resolution imagery, compared to 18.5 meters when using PlanetScope~\cite{daroya2025riverscope}. Our work aims to address this gap.

We propose \textbf{SuperRivolution}, an approach that enhances the spatial resolution of river masks by leveraging temporal information from low-resolution image sequences (\cref{fig:motivation}). To support training and evaluation of this task, we introduce a new dataset that pairs low-resolution temporal imagery with high-resolution reference labels. We evaluate a range of architectures and pretraining strategies. As a baseline, we consider a simple ensemble that aggregates predictions from individual images in a time series. Beyond this, we explore models that upsample the inputs before segmentation---either using bilinear interpolation or state-of-the-art satellite image super-resolution methods~\cite{wang2018esrgan, wolters2023zooming}. These models can be further fine-tuned on our dataset using high-resolution labels.

Our experiments show that incorporating temporal information significantly improves both segmentation accuracy and river width estimation. Moreover, training on our dataset yields consistent gains over naive aggregation approaches. For example, water segmentation improves from 60.9\% F1 score to 80.5\%, while river width estimation improves from 116.4 meters mean error to 47.8 meters. In summary, our contributions are:
\begin{itemize}
    \item A new dataset of 9,810 LR Sentinel-2 time series paired with HR ground truth labels from RiverScope~\cite{daroya2025riverscope}.
    \item We propose SuperRivolution, a temporal fusion method designed to leverage multi-image sequences to improve performance on river segmentation and width estimation.
    \item Through extensive experiments, we show that SuperRivolution significantly outperforms methods using single LR images and substantially closes the gap on models trained with expensive, HR data.
\end{itemize}
We will release the dataset, code, and pretrained models upon publication.

\section{Related Work}
\label{sec:related-work}
\noindent\textbf{Satellite image sources.}
Environmental and river monitoring has evolved from the sparse in-situ measurements to large-scale remote sensing, enabled by the vast spatiotemporal availability from satellite imagery~\cite{manfreda2024advancing}.
The increase in satellite missions over the years has resulted in a diverse set of satellite image sources~\cite{menzel1994introducing, tapley2004grace, dagras1995spot, cohen1999validating, eros2020_landsat_l2_c2, esa2022sentinel, jutz2020copernicus}.
These missions differ in several key attributes that impact remote sensing applications: temporal frequency, spatial resolution, area of coverage, and mode of access (\ie commercial or freely available)~\cite{miller2024deep}.

Public agencies like NASA and ESA provide freely accessible, near-global coverage imagery from missions like Landsat~\cite{eros2020_landsat_l2_c2} and Sentinel~\cite{esa2022sentinel}.
While the spatiotemporal coverage of these free data are invaluable, their moderate spatial resolution (10 to 30 meter/pixel) and lower temporal frequency (5 to 16 days revisit frequency) limit their effectiveness for fine-grained monitoring.
Previous works have shown that using HR imagery results in better performance due to the presence of finer spatial detail~\cite{daroya2025riverscope, mansaray2021comparing, tayer2023improving}.

To address the resolution gap, HR data sources are available, though often with compromises.
For instance, aerial programs like NAIP~\cite{usda_naip_imagery} offer free HR imagery (2 to 0.6 meter/pixel) but are geographically restricted to the United States.
Commercial vendors fill this gap by providing global, HR (1-3 meter/pixel), and high-frequency (1 to 3 days revisit frequency) multispectral imagery from satellites like PlanetScope~\cite{planetlabs}, WorldView-3~\cite{kruse2015validation}, and GeoEye-1~\cite{crespi2010geoeye}. 
However, the high operational cost associated with commercial satellites make their data prohibitively expensive for many large-scale applications.
Consequently, the selection of an appropriate data source remains a critical choice, balancing application-specific accuracy requirements against budgetary and coverage constraints.

\vspace{3pt}
\noindent\textbf{Remote sensing for river monitoring.} 
A critical step in remote river monitoring is the accurate segmentation of water bodies from satellite imagery, which enables the measurement of key hydrological parameters like width and flow~\cite{li2003remote, gardner2023human, dethier2020toward}.
Early works rely on thresholding spectral indices like the Normalized Difference Water Index (NDWI)~\cite{mcfeeters1996use}---a threshold-based method that uses the histogram of pixel values to separate land from water. 
However, these methods tend to have low accuracy especially on water boundaries~\cite{daroya2025riverscope, flores2024mapping}.

Consequently, a large body of work has emerged that explores deep learning methods for finding water bodies and rivers from satellite images~\cite{isikdogan2019seeing, flores2024mapping, daroya2025improving}.
Due to the growing availability of labeled training data, both convolutional and transformer based models have demonstrated strong performance~\cite{bastani2023satlaspretrain, jakubik2310foundation_prithvi, manas2021seasonal_seco, daroya2025wildsat}.
At the same time, pretrained segmentation architectures such as ImageNet pretrained UNet~\cite{ronneberger2015u}, FPN~\cite{long2015fully}, and DeepLabv3~\cite{chen2017deeplab} make it feasible to apply transfer learning on downstream river tasks, further increasing performance on downstream river-specific tasks~\cite{daroya2025improving}.

However, these models are still fundamentally limited by the resolution and the quality of the single input image.
Since satellite imagery are typically optical in nature, they are susceptible to atmospheric noise, such as clouds, that can occlude and affect the visibility of rivers~\cite{isikdogan2019seeing, langhorst2024global}.
At the same time, the segmentation accuracy is inherently limited by the image resolution; the error is bounded by the data's resolution (meters/pixel), making precise boundary delineation from LR sources challenging~\cite{daroya2025riverscope, mansaray2021comparing, tayer2023improving}.

\begin{figure*}[ht]
    \centering
    \includegraphics[width=0.9\linewidth]{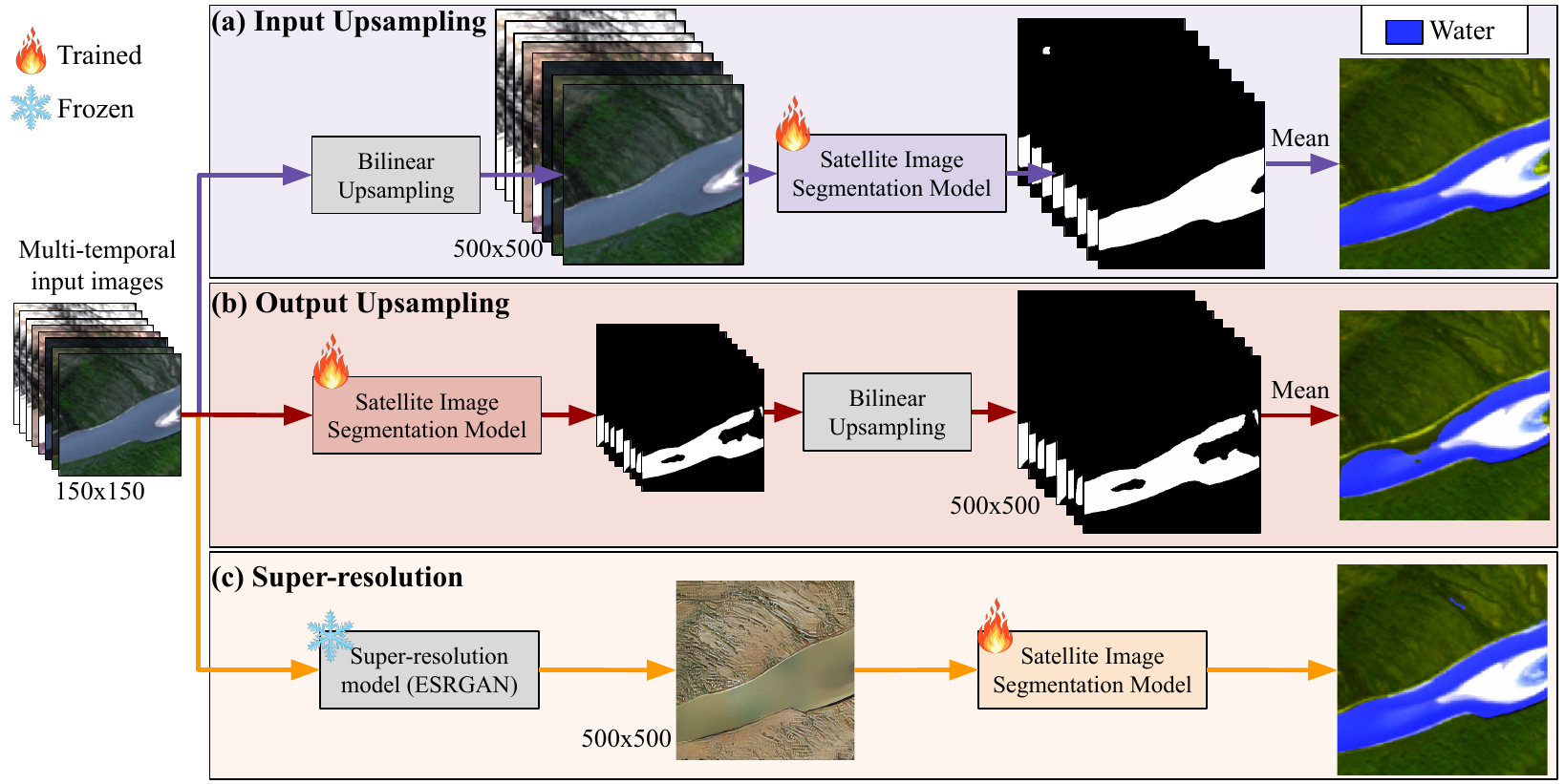}
    \vspace{-9pt}
    \caption{\textbf{Architecture for the three proposed methods}. SuperRivolution takes advantage of the availability of multiple LR satellite images in three different ways. (a) \textbf{Input upsampling} upsamples the input before feeding it to the model. (b) \textbf{Output upsampling} upsamples the LR output after applying the model to match the HR label. (c) \textbf{Super-resolution} uses a pretrained super-resolution model ESRGAN~\cite{wang2018esrgan} to get a higher resolution version of the input image.
    }
    \label{fig:architecture}
    \vspace{-9pt}
\end{figure*}

\vspace{3pt}
\noindent\textbf{Temporal fusion and super-resolution from image sequences.}
To overcome limitations of single LR images, such as occlusions from atmospheric noise (\eg clouds) and the lack of spatial detail, a common strategy is to use information from a temporal sequence of images~\cite{manfreda2024advancing}.
While single image super-resolution methods exist, they are intrinsically ill-posed and are prone to hallucinate details since they must infer high-frequency information from a single source~\cite{yu2019semantic, yu2017hallucinating, lu2022transformer}.

In contrast, multi-frame super-resolution techniques use multiple, slightly different LR frames of the same scene to construct a single HR image~\cite{li2010multi, di2025qmambabsr, dudhane2023burstormer}.
These methods are generally more reliable since they exploit the natural sub-pixel shifts between frames to recover genuine high-frequency details, leading to a more accurate reconstruction.

This technique has been successfully applied in remote sensing to generate HR imagery from a series of LR satellite images~\cite{wolters2023zooming, martens2019super, cornebise2022open, li2024learning}.
Prior work has explored various generative models including generative adversarial networks (GANs), convolutional networks, and diffusion models, finding GAN-based approaches to be the most effective~\cite{wolters2023zooming}.
Notably, the improved resolution from GAN-based super-resolution has been shown to directly enhance performance on downstream semantic segmentation tasks for industrial applications~\cite{frizza2022semantically}.
Motivated by these successes, our work applies the temporal fusion approach to improve the accuracy and robustness of downstream river monitoring tasks, specifically water segmentation and river width estimation.

\section{Method}
\label{sec:method}
\subsection{Problem Overview}
We define the problem as follows. Let $\mathcal{Y}=\{\mathbf{y}_i\}_{i=1}^N$ denote the set of HR (3~m/px) water segmentation labels, where $\mathbf{y}_i \in \{0,1\}^{W \times H}$. For each label $\mathbf{y}_i$, we have a corresponding set of temporal, LR (10~m/px) satellite images,
\begin{equation}
    \mathcal{X}_i=\{\mathbf{x}_i^{t_j}\}_{j=1}^m, \quad \mathbf{x}_i^{t_j} \in \mathbb{R}^{w \times h \times C}
    \label{eqn:low-resolution-inputs}
\end{equation}
where $\mathbf{x}_i^{t_j}$ is a $C$-channel multispectral image of location $i$ at time $t_j$. The complete dataset is $\mathcal{D} = \{ \left(\mathcal{X}_i, \mathbf{y}_i \right)\}_{i=1}^N$.

Our objective is to learn a segmentation model $f_{\theta}$ that predicts an HR segmentation mask $\hat{\mathbf{y}}_i = f_{\theta}(\mathcal{X}_i)$ from the LR image sequence.
The success of our model is measured by its ability to close the performance gap between models trained on LR versus HR data.
We evaluate $f_{\theta}$ against (1) an upper-bound oracle, and (2) an LR baseline.
The oracle model is trained on single HR images and their corresponding HR labels $\mathbf{y}_i$.
Our goal is to approach the performance of this oracle without access to HR imagery at inference time.
At the same time, we aim to outperform an LR baseline trained on single LR images paired with LR labels.

Performance is assessed on downstream river monitoring tasks, specifically water segmentation and river width estimation.
The river widths are derived directly from the predicted water masks using the proposed algorithm in~\cite{yang2019rivwidthcloud}.
This algorithm calculates width by counting the water pixels along the intersections perpendicular to the river centerline at predefined locations, and multiplying this number by the image spatial resolution.

\subsection{SuperRivolution Architecture}
We explore three methods for predicting HR water masks from multiple LR images: input upsampling, output upsampling, and super-resolution.
The input and output upsampling methods apply the same model across multiple images, and aggregates the predictions.
Super-resolution uses a pretrained super-resolution model on the input images. 
\cref{fig:architecture} shows the different methods.

For each of these three methods, different segmentation models are used covering different ways of extracting features: UNet~\cite{ronneberger2015u}, FPN~\cite{long2015fully}, DeepLabv3 (DLv3)~\cite{chen2017deeplab}, and DPT~\cite{ranftl2021vision}. 
DPT is applied to ViT-based backbones, while the others are applied to convolutional- and Swin-based backbones.
Backbones include ResNet50 (RN50)~\cite{heDeepResidualLearning2016}, MobileNetv2 (MV2)~\cite{sandler2018mobilenetv2}, Swin-T~\cite{liu2021swin}, Swin-B~\cite{liu2021swin}, ViT-B/16~\cite{dosovitskiy2020vit}, and ViT-L/16~\cite{dosovitskiy2020vit}.
We also explore different pretraining methods such as supervised methods using SatlasNet~\cite{bastani2023satlaspretrain} and ImagetNet1k~\cite{deng2009imagenet}, and self-supervised methods using SeCo~\cite{manas2021seasonal_seco}, MoCov3~\cite{chen2021empirical_mocov3}, CLIP~\cite{radford2021learning_clip}, and DINO~\cite{caron2021emerging}.
Among these, SeCo and SatlasNet use satellite images for pretraining.
We use combinations that have open-source pretrained weights.\\

\noindent\textbf{Input Upsampling} upsamples the LR input image to match the dimensions of the HR label \emph{before} applying the model $f_{\theta}$.
For example, to align a $500 \times 500$ pixel target label $\mathbf{y}_i$ at 3~m/px, a corresponding $150 \times 150 \times C$ LR input $\mathbf{x}_i^{t_j}$ at 10~m/px is first bilinearly upsampled to $\mathrm{Up}\left(\mathbf{x}_i^{t_j}\right) \in \mathbb{R}^{500 \times 500 \times C}$ that is then fed into the model.
When multiple ($m$) LR images of the same location are available, this process is applied to each image independently using the same model $f_{\theta}$, which is trained to produce a single output from a single image.
The final prediction is then calculated by averaging the logits across all $m$ outputs.
The prediction $\hat{\mathbf{y}}_i$ is computed as:
\begin{equation}
    \hat{\mathbf{y}}_i =  \frac{1}{m} \sum_{j=1}^m f_{\theta} \left( \mathrm{Up} \left(\mathbf{x}_i^{t_j}\right) \right).
    \label{eqn:input-upsampling}
\end{equation}

\noindent\textbf{Output Upsampling} bilinearly upsamples the image \emph{after} applying the model $f_{\theta}$ when fed with the LR input image $\mathbf{x}_i^{t_j}$. Similar to the input upsampling method, when $m$ LR images are available, $f_{\theta}$ is applied to the images independently and the final output is the average prediction (logits).
$f_{\theta}$ is trained to produce a single output from a single image.
The prediction can be computed as:

\begin{equation}
    \hat{\mathbf{y}}_i =  \frac{1}{m} \sum_{j=1}^m \mathrm{Up} \left( f_{\theta} \left( \mathbf{x}_i^{t_j} \right) \right).
    \label{eqn:output-upsampling}
\end{equation}

\noindent\textbf{Super-resolution} uses a pre-trained super-resolution model $g_{\phi}$ to produce an HR image from single or multiple LR images. We use ESRGAN~\cite{wang2018esrgan} as $g_{\phi}$ since it was shown to outperform other methods on satellite image super-resolution~\cite{wolters2023zooming}. The prediction can be represented as:

\begin{equation}
    \hat{\mathbf{y}}_i = f_{\theta} \big( g_{\phi} \left( \mathcal{X}_i \right) \big).
    \label{eqn:super-resolution}
\end{equation}

\subsection{Training}
To leverage features from models pretrained on RGB datasets like ImageNet, we use a linear adaptor on our multispectral input.
This adaptor linearly projects the C-channel input to 3 channels, a strategy shown to be more effective than training from scratch or dropping spectral channels~\cite{daroya2025riverscope}.

Following the best setting observed by previous multi-frame super-resolution works, all temporal models use a fixed-length sequence of 8 images as input~\cite{wolters2023zooming}.
When more than 8 frames are available for a given location, we select the 8 frames with the least number of no-data pixels to ensure input quality.
All models were trained for 50 epochs using the Adam optimizer.
We use the binary cross-entropy loss for training, defined as:

\begin{multline}
    \mathcal{L}_{bce}\left(\mathbf{y}_i, \hat{\mathbf{y}}_i \right) = \frac{1}{WH} \sum_{j=1}^W \sum_{k=1}^H - \Biggl(y_{i,(j,k)} \log \hat{y}_{i,(j,k)} \\+ \left(1-y_{i,(j,k)} \right) \log\left(1-\hat{y}_{i,(j,k)} \right) \Biggr).
    \label{eq:bce}
\end{multline}

For each configuration, the learning rate and the final model checkpoint were selected based on the best segmentation performance on a held-out validation set.
All experiments were conducted on a single NVIDIA L40S GPU.

\section{Experiments}
\subsection{Dataset}
\label{subsec:dataset}
Our work builds on the RiverScope dataset~\cite{daroya2025riverscope}, a public collection of 1,145 high-resolution (3~m/px) multispectral satellite images from PlanetScope.
Each image is paired with high-fidelity ground truth labels for water segmentation and river width estimation.
We select this dataset over alternatives with lower-resolution labels such as WorldCover~\cite{van2021esa_worldcover, zanaga2022esa}, which can limit model performance due to imprecise boundary definitions~\cite{daroya2025riverscope}.
To ensure fair comparison with existing benchmarks, we adopt the original training, validation, and test splits defined in RiverScope.

We augment RiverScope with LR time-series imagery from Sentinel-2~\cite{esa2022sentinel}.
For each HR label, we collect a corresponding sequence of 10~m/px multispectral Sentinel-2 images from the same geographic location.
Centered on the timestamp of the original RiverScope image, we sample at least 8 images from a four-month window ($\pm$ 2 months), taking into account Sentinel's up to 2 week revisit frequency.
To have proper spatial alignment, all Sentinel-2 images are reprojected and co-registered to match the projection and boundaries of the RiverScope labels.
This process resulted in a new dataset of 9,810 Sentinel-2 images, each with 12 bands.
We will release both the raw and the reprojected images for future research.

\subsection{Evaluation}
We evaluate performance of all methods using the held-out test set defined in RiverScope.
Performance is measured on two downstream river monitoring tasks: (1) water segmentation, and (2) river width estimation.

Water segmentation is a direct output of the models.
The performance is measured through F1 score, precision, and recall.
F1 score is used due to the imbalance in the ratio of water pixels to non-water pixels (there is an average of 20\% water pixels in an image).

We estimate river width by adopting the methodology from RiverScope.
At predefined locations, the width is computed by multiplying the number of water pixels perpendicular to the river's centerline by the image's spatial resolution~\cite{yang2019rivwidthcloud}.
We calculate bias, \% bias, mean absolute error, and median absolute error to capture the difference in performance.
The bias is computed as the average of the difference between the predicted widths $\hat{y}_i$ and the ground truth widths $y_i$:
\begin{equation}
    \text{bias} = (\frac{1}{N} \sum_{i=1}^N \left( \hat{y}_i-y_i \right)).
\end{equation}
The \% bias divides each prediction with the ground truth width: 
\begin{equation}
    \text{\% bias} = \frac{1}{N} \sum_{i=1}^N \frac{\hat{y}_i - y_i}{y_i} \times 100\%.   
\end{equation}
The mean and median absolute errors are calculated as the mean and median of $|y_i - \hat{y}_i| \ \forall i$, respectively.

\subsection{Baselines}
To better evaluate our proposed method, SuperRivolution, we compare it against two key baselines.
These are designed to isolate the specific advantages of our approach, which combines multiple LR inputs with a single HR label.
For a fair comparison, all models share the same architecture and training configurations.

\vspace{3pt}

\noindent\textbf{Sentinel Baseline (Low-resolution baseline).} This model establishes a performance lower-bound using a standard, LR setup. 
It is trained on 50,000 LR Sentinel-2 images paired with corresponding LR labels from WorldCover. 
This baseline demonstrates the performance achievable without access to any HR training data.
To evaluate its predictions against our HR ground truth, its output is upsampled using bilinear interpolation.
When using a temporal input, we also take the mean of the predictions, similar to the output upsampling method of SuperRivolution~(\cref{eqn:output-upsampling}).

\vspace{3pt}

\noindent\textbf{RiverScope (High-resolution baseline).} This model acts as an \emph{upper-bound} benchmark, showing the performance possible with ideal, HR data.
It is trained on HR PlanetScope images and their corresponding HR labels. 
The goal of SuperRivolution is to bridge the performance gap between this oracle and the LR baseline, using only LR images at inference time.

\section{Results and Discussion}
\label{sec:experiments}

\begin{figure*}[ht]
    \centering
    \includegraphics[width=0.9\linewidth]{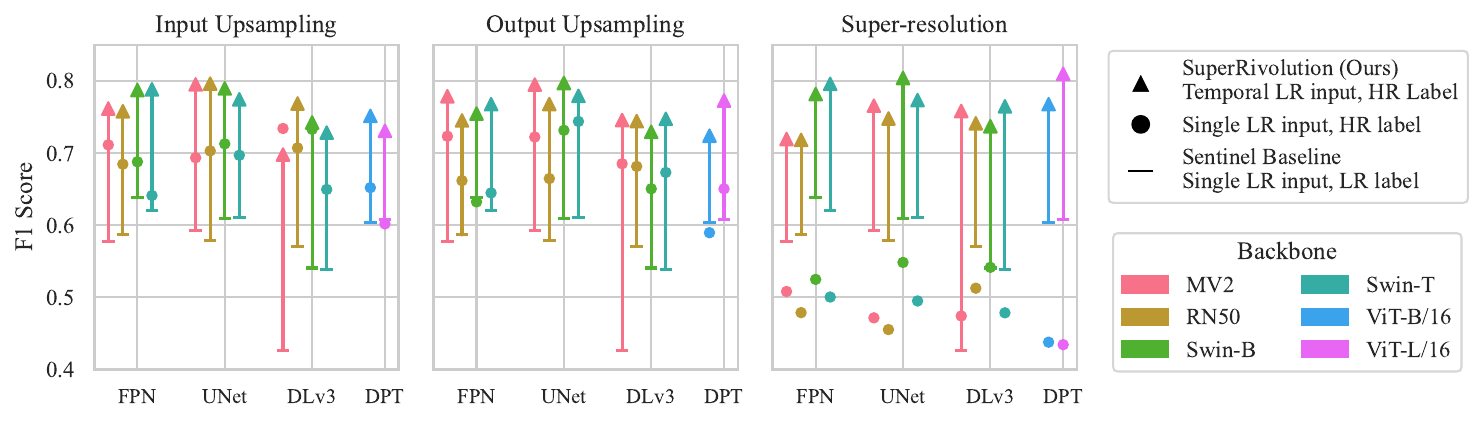}
    \vspace{-12pt}
    \caption{\textbf{Segmentation performance across different methods when compared with low-resolution baseline}. Each entry is the average performance across different pretraining methods. For all methods across backbones and segmentation models, the segmentation performance significantly improved when using temporal SuperRivolution. We include results across pretraining methods in the Appendix.
    }
    \label{fig:segmentation-metrics}
    \vspace{-12pt}
\end{figure*}

\begin{table*}[t]
    \centering
    \small
    \setlength{\tabcolsep}{4pt}
    
    \begin{tabular}{@{}llccc@{}}
    \toprule
    \textbf{Group} & \textbf{Method} & \textbf{F1 Score} & \textbf{Recall} & \textbf{Precision} \\
    \midrule
    \multicolumn{5}{@{}l}{\textit{\textbf{Baselines}}} \\
     \hspace{3mm}Upper Bound (HR) & RiverScope~\cite{daroya2025riverscope} & $94.1 \pm 0.6$ & $95.2 \pm 0.7$ & $93.0 \pm 1.8$ \\
     \hspace{3mm}Lower Bound (LR) & Sentinel Baseline & $60.9 \pm 0.6$ & $94.3 \pm 0.9$ & $45.0 \pm 0.5$ \\
    \midrule
    \multicolumn{5}{@{}l}{\textbf{SuperRivolution (Ours, Temporal LR)}} \\
     & Input Upsampling & $79.2 \pm 0.6$ & $83.6 \pm 5.7$ & $75.6 \pm 3.3$ \\
     & Output Upsampling & $79.4 \pm 1.3$ & $83.2 \pm 5.8$ & $76.2 \pm 3.5$ \\
     & Super-resolution & $\mathbf{80.5 \pm 1.5}$ & $77.8 \pm 2.4$ & $83.8 \pm 5.5$ \\
    \bottomrule
    \end{tabular}

    \vspace{-6pt}
    \caption{\textbf{Quantitative comparison of segmentation performance}. Results show mean $\pm$ standard deviation (\%) for each method on the test set using a UNet Swin-B pretrained on ImageNet1k. LR and HR denote low (10~m/px) and high (3~m/px) resolution, respectively. Temporal SuperRivolution methods bridge the performance gap between the LR Sentinel baseline and the HR RiverScope upper-bound.
    }
    \label{tab:segmentation-performance_comparison}
    \vspace{-6pt}
\end{table*}

\begin{table}[t]
    \centering
    \small
    \setlength{\tabcolsep}{2pt}
    \begin{tabular}{@{}llccc@{}}
    \toprule
    \textbf{Method} & \textbf{Input Type} & \textbf{F1 Score} & \textbf{Recall} & \textbf{Precision} \\
    \midrule
    \multicolumn{5}{@{}l}{\textit{\textbf{Baseline}}} \\
    \hspace{3mm}Sentinel Baseline & Single LR & $60.9$ & $94.3$ & $45.0$ \\
     & Temporal LR & $75.9$ & $71.8$ & $81.1$ \\
    \midrule
    \multicolumn{5}{@{}l}{\textbf{SuperRivolution (Ours)}} \\
    \hspace{3mm}Input Upsampling & Single LR & $71.7$ & $63.7$ & $82.2$ \\
     & Temporal LR & $79.2$ & $83.6$ & $75.6$ \\
    \cmidrule(l){2-5}
    \hspace{3mm}Output Upsampling & Single LR & $73.2$ & $68.0$ & $80.1$ \\
     & Temporal LR & $79.4$ & $83.2$ & $76.2$ \\
    \cmidrule(l){2-5}
    \hspace{3mm}Super-resolution & Single LR & $53.9$ & $59.0$ & $49.7$ \\
     & Temporal LR & $\mathbf{80.5}$ & $77.8$ & $83.8$ \\
    \bottomrule
    \end{tabular}
    \vspace{-6pt}
    \caption{\textbf{Impact of temporal information on segmentation performance}. Shifting from a single LR input to a temporal LR input shows consistent positive gain in F1 score.
    }
    \label{tab:segmentation-performance-ablation}
    \vspace{-9pt}
\end{table}

\subsection{Water Segmentation}

\noindent\textbf{Using temporal LR images results in better segmentation performance.}
\cref{fig:segmentation-metrics} shows that using multiple LR images can ultimately help boost segmentation performance.
For instance, a UNet Swin-B improves to 79.2\% F1 score from 60.9\% using input upsampling.
In the absence of HR satellite images due to cost constraints, the availability of multiple LR images can be a promising alternative.
Although using a single LR image paired with an HR label can sometimes improve performance (71.7\% F1 score for UNet Swin-B input upsampling), it is unreliable compared to using temporal images.
In \cref{fig:segmentation-metrics}, some of the single LR input models end up performing worse than the Sentinel baseline (\eg DPT ViT-B/16 F1 score decreases to 43.8\% from 60.4\% using super-resolution with single LR).
\cref{tab:segmentation-performance_comparison} and \cref{tab:segmentation-performance-ablation} further show the exact performance improvement for each change.
Notably, our best model (80.5\% F1 score) effectively bridges the gap between the LR Sentinel baseline (60.9\%) and the HR RiverScope upper-bound (94.1\%).

\noindent\textbf{Aligning HR labels with single low-resolution satellite images is insufficient for optimal performance.}
\cref{tab:segmentation-performance-ablation} shows using temporal LR inputs yield far greater improvements, with as much as a 26.6\% F1 score improvement from single to temporal LR for super-resolution.
A single image approach is more susceptible to atmospheric noise, such as cloud cover, which can degrade recall, especially if the label is paired with a cloudy image.
While HR labels can improve precision by sharpening boundary definitions and reducing false positives, the gain is often offset by the drop in recall when using single image inputs.
In contrast, a temporal approach mitigates the impact of poor quality inputs, improving precision without sacrificing recall.

\begin{figure}[!t]
    \centering
    \includegraphics[width=0.95\linewidth]{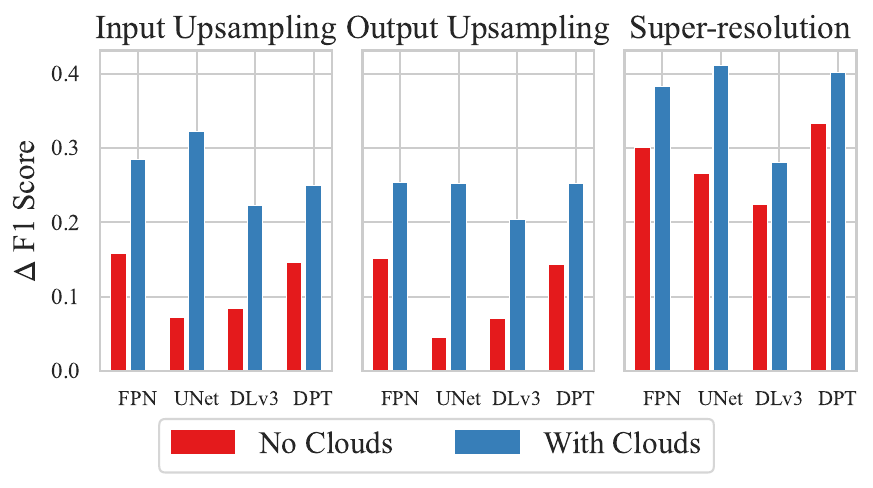}
    \vspace{-12pt}
    \caption{\textbf{Impact of cloud cover on temporal input performance gain}. The plot shows the gain in F1 Score ($\Delta F1$) when using multiple images instead of one, under varying cloud conditions. For clarity, only ViT and Swin backbones are shown, but trends are consistent across configurations (see Appendix).
    }
    \label{fig:segmentation-cloud-effect}
    \vspace{-12pt}
\end{figure}

\begin{figure*}[!t]
    \centering
    \includegraphics[width=0.9\linewidth]{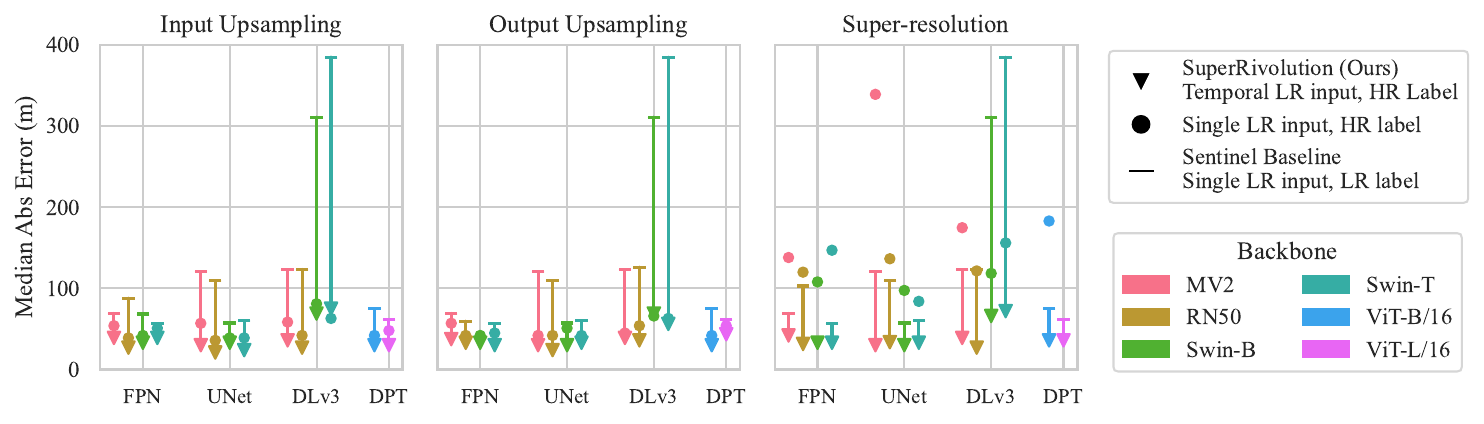}
    \vspace{-12pt}
    \caption{\textbf{River width estimation error across different methods when compared with the LR baseline}. Across all settings, using temporal SuperRivolution results in lower error. We include results across pretraining methods in the Appendix.
    }
    \label{fig:river-width-metrics}
    \vspace{-15pt}
\end{figure*}

\noindent\textbf{Using multiple images is more robust to cloud cover.} 
\cref{fig:segmentation-cloud-effect} shows the visualization of performance improvements from using single image inputs to temporal inputs under varying cloud conditions.
The presence of clouds in each Sentinel-2 tile was determined using the cloud cover metadata approximated using the Fmask algorithm~\cite{qiu2019fmask}.
Images with at least 10\% cloud cover were identified as cloudy.
Results show that the F1 score improvement going from single to multiple images is higher when the images are cloudy (\eg as much as 40\% F1 score improvement on cloudy images when using super-resolution).
This explains that clouds have a significant effect on performance and can be mitigated with multiple image inputs.

\noindent\textbf{Super-resolution models benefit the most from multiple images.} 
\cref{fig:segmentation-cloud-effect} shows that super-resolution models improve significantly from single image inputs to multiple image inputs, regardless of cloud conditions.
The F1 score improvement on non-cloudy images for super-resolution can sometimes even be higher than the improvement for cloudy images for the input and output upsampling methods.
For instance, there is a 30\% F1 score improvement for FPN using super-resolution on non-cloudy images, while FPN using output upsampling on cloudy images results in 25\% improvement.
Apart from the advantage of being less sensitive to clouds, using multiple images also result in better super-resolution performance~\cite{wolters2023zooming}.
Thus, the super-resolution methods see the most gains with the temporal setup, and also have the higher F1 score for segmentation among the other methods for temporal LR input (\cref{tab:segmentation-performance_comparison}).

\begin{table}[t]
    \centering
    \small
    \setlength{\tabcolsep}{3pt}
    \begin{tabular}{@{}lcccc@{}}
    \toprule
    \textbf{Method} & \textbf{Mean} & \textbf{Median} & \textbf{Bias} & \textbf{\% Bias} \\
    \midrule
    \multicolumn{5}{@{}l}{\textit{\textbf{Baselines}}} \\
    \hspace{3mm}RiverScope~\cite{daroya2025riverscope} & 18.5 & 12.0 & 8.9 & 16.8\% \\
    \hspace{3mm}Sentinel Baseline & 116.4 & 62.0 & 93.9 & 91.2\% \\
    \midrule
    \multicolumn{5}{@{}l}{\textbf{SuperRivolution (Ours, Temporal LR)}} \\
    \hspace{3mm}Input Upsampling & \textbf{47.8} & 33.0 & 7.0 & 40.5\% \\
    \hspace{3mm}Output Upsampling & 53.0 & \textbf{30.0} & -12.2 & \textbf{20.1\%} \\
    \hspace{3mm}Super-resolution & 50.2 & \textbf{30.0} & \textbf{2.0} & 26.5\% \\
    \bottomrule
    \end{tabular}%
    \vspace{-9pt}
    \caption{\textbf{River width estimation errors (meters) of different methods}. The absolute errors are used for mean and median. All methods use UNet Swin-B pretrained with ImageNet1k for a direct comparison. RiverScope is the upper-bound benchmark.}
    \label{tab:width-comparison}
    \vspace{-12pt}
\end{table}

\begin{table}[t]
    \centering
    \small
    \setlength{\tabcolsep}{2pt}
    \begin{tabular}{@{}llcc@{}}
    \toprule
    \textbf{Method} & \textbf{Input Type} & \textbf{Mean} & \textbf{Median}\\
    \midrule
    \multicolumn{4}{@{}l}{\textit{\textbf{Baseline}}} \\
    {\hspace{3mm}Sentinel Baseline} & Single LR & 116.4  & 62.0  \\
     & Temporal LR & 92.4  & 50.0 \\
    \midrule
    \multicolumn{4}{@{}l}{\textbf{SuperRivolution (Ours)}} \\
    {\hspace{3mm}Input Upsampling} & Single LR & 63.2  & 39.0 \\
     & Temporal LR & \textbf{47.8} & 33.0 \\
    \cmidrule(l){2-4}
    {\hspace{3mm}Output Upsampling} & Single LR & 86.5  & 51.0   \\
     & Temporal LR &  53.0 & \textbf{30.0}  \\
    \cmidrule(l){2-4}
    {\hspace{3mm}Super-resolution} & Single LR & 236.3  & 97.5   \\
     & Temporal LR & 50.2  & \textbf{30.0}  \\
    \bottomrule
    \end{tabular}
    
    \vspace{-9pt}
    \caption{\textbf{Impact of temporal information on width error (m)}. Using temporal LR significantly reduces river width errors.}
    \label{tab:width-comparison-ablation-temporal}
    \vspace{-15pt}
\end{table}

\subsection{River Width Estimation}

\noindent\textbf{Improved segmentation performance also resulted in improved river width estimation.}
\cref{fig:river-width-metrics} shows that the median absolute error (meters) went down across the board when temporal inputs are used.
This can also be observed in \cref{tab:width-comparison} and \cref{tab:width-comparison-ablation-temporal}.
Our best model reduces the mean error to 47.8~m from 116.4~m for the LR baseline (18.5~m for HR upper bound RiverScope).
\cref{tab:width-comparison-ablation-temporal} also shows that unlike single LR inputs, using temporal LR inputs show consistently better performance with less width errors. 

\noindent\textbf{Input upsampling is the most reliable method for river width estimation from temporal inputs.}
\cref{tab:width-comparison} and \cref{tab:width-comparison-ablation-temporal} show that of all methods with LR input, input upsampling achieves the lowest mean absolute error of 47.8~m (with comparable median error as other SuperRivolution methods).
Although the super-resolution method obtains slightly better segmentation scores (\cref{tab:segmentation-performance_comparison}), we attribute the better width estimation performance of input upsampling to its higher recall.
This suggests a more stable and reliable prediction, which is critical for accurate width measurement.

\begin{figure}[!t]
    \centering
    \includegraphics[width=0.96\linewidth]{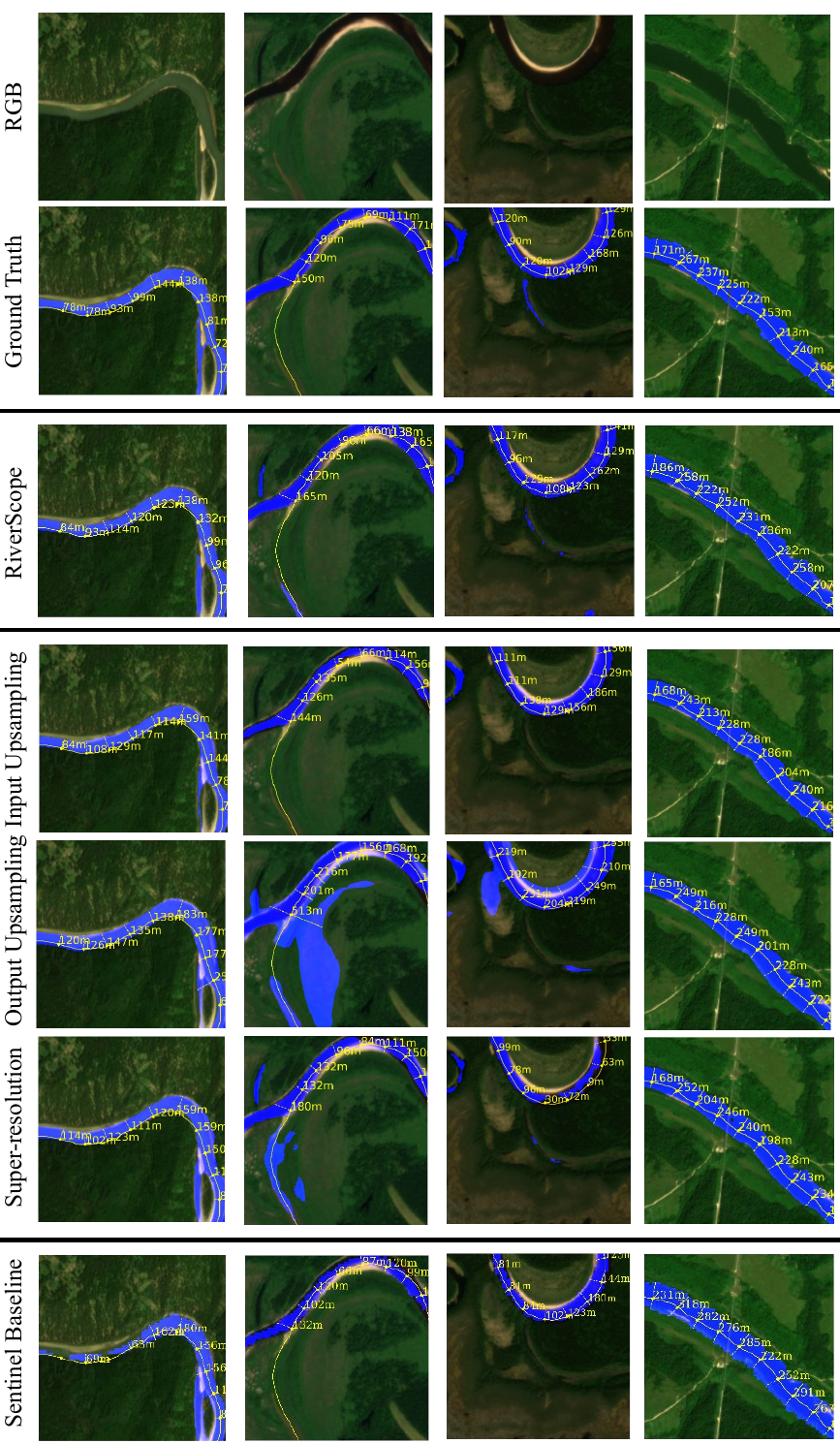}
    \vspace{-10pt}
    \caption{\textbf{Visualization of predicted river segmentation and widths across different methods}. All methods except the upper bound RiverScope use temporal inputs. Input upsampling has more reliable segmentations and width estimates compared to the other methods. We include more visualizations in the Appendix.
    }
    \label{fig:segmentation-width-viz}
    \vspace{-12pt}
\end{figure}

\cref{fig:segmentation-width-viz} shows a visualization of the segmentation and width estimates from the different methods.
Input upsampling is likely effective because it allows the model to operate directly on the HR label. This encourages the model to learn how to extract high-frequency details and corrections by directly optimizing against the HR labels.
In contrast, output upsampling confines the model to the LR domain, where the final upsampling is a post-processing step that occurs after the model has already made its prediction.
Similarly, while a super-resolution module can create an HR image, it risks introducing hallucinated artifacts which can result in errors in the final segmentation.
More examples are available in \cref{fig:supp-segmentation-viz-wbaselines} (Appendix).
Consistent with \cref{tab:width-comparison}, input upsampling more closely resembles the ground truth.

\noindent\textbf{The proposed temporal methods show reduction in prediction bias compared to baselines.}  
\cref{tab:width-comparison} shows the temporal methods show a stronger agreement with the ground truth than the Sentinel baseline, achieving lower bias metrics.
The Sentinel baseline tends to have more skewed predictions, resulting in a higher bias (see \cref{fig:supp-river-width-scatter} (Appendix)).
Furthermore, the biases of the input upsampling and super-resolution temporal models (7~m and 2~m, respectively) are less than the HR baseline RiverScope bias (8.9~m). 
This might be due to the temporal averaging present in the temporal models.
By integrating information over time, the models learn to ignore transient artifacts (\eg clouds, shadows) that can skew predictions.

\section{Limitations and Future Work}
\label{sec:limitations}
The primary limitations of our work stem from the RiverScope dataset and our temporal data fusion approach.
While RiverScope offers high label quality, its size is modest compared to global coverage datasets, a trade-off we made to prioritize fidelity of HR ground truth over the sheer volume of training data.
Furthermore, our four-month temporal window is well-suited for capturing gradual river changes like meander migration (\ie when rivers slowly shift across a floodplain), but is insufficient for monitoring rapid, event-driven phenomena such as flash floods.
However, this temporal limitation can be overcome by applying our fusion methodology to LR satellite imagery with higher temporal frequency (\eg daily acquisitions) which would enable the monitoring of more dynamic events.

Despite these constraints, our findings establish a clear trajectory for future research by showing that using temporal LR images can serve as an effective proxy when HR satellite images are limited. Future work can apply our temporal fusion methodology to more geographically diverse and larger-scale datasets to test its global robustness. To further enhance performance, future work can explore using a student-teacher distillation framework, where an accurate HR teacher model trains an LR student model, reducing the dependence on high-resolution data with human labels.

\section{Conclusion}
\label{sec:conclusion}
In this work, we addressed the challenge of high-fidelity river monitoring when high-resolution (HR) data is limited.
We contributed a new dataset of temporal low-resolution (LR) images paired with high-resolution labels, and introduced SuperRivolution, a method that leverages multiple LR satellite images to produce more accurate river segmentations and width estimates. 
Our method significantly outperforms standard models that take single LR satellite images as input and closes the gap on expensive HR models.
Our method achieved river segmentation F1 score of 80.5\% and mean width error of 47.8~m, large improvements compared to the standard model with 60.9\% F1 score and 116.4~m mean width error.
SuperRivolution presents an effective and scalable alternative for large-scale river monitoring.

Our findings show that temporal data can serve as a powerful alternative to spatial resolution in remote sensing. This opens up research that shifts the focus from acquiring scarce, expensive HR imagery toward better utilization of the large public collection of LR satellite data. We also found that even straightforward fusion techniques, such as input upsampling, provide a significant performance boost.
Crucially, such methods can be readily integrated into existing model architectures, offering a practical way to improve performance whenever temporal data is available.
As these sources of imagery continue to grow with new satellite missions, integrating them with other data sources offers a promising path toward greater improvements for river and environmental monitoring.

\paragraph{Acknowledgements.}
Experiments were performed on the University of Massachusetts GPU cluster funded by the Mass.~Technology Collaborative. RD and SM were supported in part by NASA grant 80NSSC22K1487 and NSF grant 2329927.

{
    \small
    \bibliographystyle{ieeenat_fullname}
    \bibliography{main}

@String(CVPR= {IEEE Conf. Comput. Vis. Pattern Recog.})

@String(ECCV= {Eur. Conf. Comput. Vis.})

@String(ICLR = {Int. Conf. Learn. Represent.})

@String(CVPR  = {CVPR})

@String(ECCV  = {ECCV})

@String(ICLR  = {ICLR})

@inproceedings{li2024learning,
  title={Learning without exact guidance: Updating large-scale high-resolution land cover maps from low-resolution historical labels},
  author={Li, Zhuohong and He, Wei and Li, Jiepan and Lu, Fangxiao and Zhang, Hongyan},
  booktitle={Proceedings of the IEEE/CVF Conference on Computer Vision and Pattern Recognition},
  pages={27717--27727},
  year={2024}
}

@misc{usda_naip_imagery,
  author       = {{USDA Farm Service Agency}},
  title        = {NAIP Imagery},
  year         = {n.d.},
  howpublished = {\url{https://www.fsa.usda.gov/programs-and-services/aerial-photography/imagery-programs/naip-imagery/}},
}

@inproceedings{van2021esa_worldcover,
  title={ESA WorldCover: Global land cover mapping at 10 m resolution for 2020 based on Sentinel-1 and 2 data.},
  author={Van De Kerchove, Ruben and Zanaga, Daniele and Keersmaecker, Wanda and Souverijns, Niels and Wevers, Jan and Brockmann, Carsten and Grosu, Alex and Paccini, Audrey and Cartus, Oliver and Santoro, Maurizio and others},
  booktitle={AGU Fall Meeting Abstracts},
  volume={2021},
  pages={GC45I--0915},
  year={2021}
}

@article{esa2022sentinel,
  title={Sentinel-1-missions-Sentinel online-Sentinel online},
  author={ESA},
  journal={Eur. Sp. Agency},
  year={2022}
}

@dataset{eros2020_landsat_l2_c2,
  author    = {{Earth Resources Observation and Science (EROS) Center}},
  title     = {Landsat 8–9 Operational Land Imager / Thermal Infrared Sensor Level-2, Collection 2},
  year      = {2020},
  publisher = {U.S. Geological Survey},
  doi       = {10.5066/P9OGBGM6},
  url       = {https://doi.org/10.5066/P9OGBGM6},
  note      = {Dataset}
}

@article{mansaray2021comparing,
  title={Comparing PlanetScope to Landsat-8 and Sentinel-2 for sensing water quality in reservoirs in agricultural watersheds},
  author={Mansaray, Abubakarr S and Dzialowski, Andrew R and Martin, Meghan E and Wagner, Kevin L and Gholizadeh, Hamed and Stoodley, Scott H},
  journal={Remote Sensing},
  volume={13},
  number={9},
  pages={1847},
  year={2021},
  publisher={MDPI}
}

@article{tayer2023improving,
  title={Improving the accuracy of the Water Detect algorithm using Sentinel-2, Planetscope and sharpened imagery: a case study in an intermittent river},
  author={Tayer, Thiaggo C and Douglas, Michael M and Cordeiro, Maur{\'\i}cio CR and Tayer, Andr{\'e} DN and Callow, J Nik and Beesley, Leah and McFarlane, Don},
  journal={GIScience \& Remote Sensing},
  volume={60},
  number={1},
  pages={2168676},
  year={2023},
  publisher={Taylor \& Francis}
}

@article{daroya2025riverscope,
  title={RiverScope: High-Resolution River Masking Dataset},
  author={Rangel Daroya and Taylor Rowley and Jonathan Flores and Elisa Friedmann and Fiona Bennitt and Heejin An and Travis Simmons and Marissa Jean Hughes and Camryn L Kluetmeier and Solomon Kica and J. Daniel Vélez and Sarah E. Esenther and Thomas E. Howard and Yanqi Ye and Audrey Turcotte and Colin Gleason and Subhransu Maji},
  journal={arXiv preprint arXiv:2509.02451},
  year={2025}
}

@article{wolters2023zooming,
  title={Zooming out on zooming in: Advancing super-resolution for remote sensing},
  author={Wolters, Piper and Bastani, Favyen and Kembhavi, Aniruddha},
  journal={arXiv preprint arXiv:2311.18082},
  year={2023}
}

@article{jutz2020copernicus,
  title={Copernicus: the european earth observation programme},
  author={Jutz, S and Milagro-Perez, M Pilar},
  journal={Revista de Teledetecci{\'o}n},
  number={56},
  pages={V--XI},
  year={2020}
}

@article{menzel1994introducing,
  title={Introducing GOES-I: The first of a new generation of geostationary operational environmental satellites},
  author={Menzel, W Paul and Purdom, James FW},
  journal={Bulletin of the American Meteorological Society},
  volume={75},
  number={5},
  pages={757--782},
  year={1994},
  publisher={American Meteorological Society}
}

@article{tapley2004grace,
  title={GRACE measurements of mass variability in the Earth system},
  author={Tapley, Byron D and Bettadpur, Srinivas and Ries, John C and Thompson, Paul F and Watkins, Michael M},
  journal={Science},
  volume={305},
  number={5683},
  pages={503--505},
  year={2004},
  publisher={American Association for the Advancement of Science}
}

@article{dagras1995spot,
  title={The SPOT-5 mission},
  author={Dagras, CH and Duran, M and Zarrouati, Olivier and Fratter, C},
  journal={Acta Astronautica},
  volume={35},
  number={9-11},
  pages={651--660},
  year={1995},
  publisher={Elsevier}
}

@article{miller2024deep,
  title={Deep learning for satellite image time-series analysis: A review},
  author={Miller, Lynn and Pelletier, Charlotte and Webb, Geoffrey I},
  journal={IEEE Geoscience and Remote Sensing Magazine},
  volume={12},
  number={3},
  pages={81--124},
  year={2024},
  publisher={IEEE}
}

@misc{planetlabs,
  author =    {Planet Labs PBC},
  organization = {Planet},
  title =     {Planet Application Program Interface: In Space for Life on Earth},
  year =      {2024},
  url = "https://api.planet.com"
}

@article{cohen1999validating,
  title={Validating MODIS terrestrial ecology products: linking in situ and satellite measurements},
  author={Cohen, Warren B and Justice, Christopher O},
  journal={Remote Sensing of Environment},
  volume={70},
  number={1},
  pages={1--3},
  year={1999}
}

@article{daroya2025improving,
  title={Improving Satellite Imagery Masking using Multi-task and Transfer Learning},
  author={Daroya, Rangel and Lucchese, Luisa Vieira and Simmons, Travis and Prum, Punwath and Pavelsky, Tamlin and Gardner, John and Gleason, Colin J and Maji, Subhransu},
  journal={IEEE Journal of Selected Topics in Applied Earth Observations and Remote Sensing},
  year={2025},
  publisher={IEEE}
}

@article{mcfeeters1996use,
  title={The use of the Normalized Difference Water Index (NDWI) in the delineation of open water features},
  author={McFeeters, Stuart K},
  journal={International journal of remote sensing},
  volume={17},
  number={7},
  pages={1425--1432},
  year={1996},
  publisher={Taylor \& Francis}
}

@article{flores2024mapping,
  title={Mapping proglacial headwater streams in High Mountain Asia using PlanetScope imagery},
  author={Flores, Jonathan A and Gleason, Colin J and Brinkerhoff, Craig B and Harlan, Merritt E and Lummus, M Malisse and Stearns, Leigh A and Feng, Dongmei},
  journal={Remote Sensing of Environment},
  volume={306},
  pages={114124},
  year={2024},
  publisher={Elsevier}
}

@article{isikdogan2019seeing,
  title={Seeing through the clouds with deepwatermap},
  author={Isikdogan, Leo F and Bovik, Alan and Passalacqua, Paola},
  journal={IEEE Geoscience and Remote Sensing Letters},
  volume={17},
  number={10},
  pages={1662--1666},
  year={2019},
  publisher={IEEE}
}

@article{frizza2022semantically,
  title={Semantically accurate super-resolution generative adversarial networks},
  author={Frizza, Tristan and Dansereau, Donald G and Seresht, Nagita Mehr and Bewley, Michael},
  journal={Computer Vision and Image Understanding},
  volume={221},
  pages={103464},
  year={2022},
  publisher={Elsevier}
}

@inproceedings{ronneberger2015u,
  title={U-net: Convolutional networks for biomedical image segmentation},
  author={Ronneberger, Olaf and Fischer, Philipp and Brox, Thomas},
  booktitle={Medical image computing and computer-assisted intervention--MICCAI 2015: 18th international conference, Munich, Germany, October 5-9, 2015, proceedings, part III 18},
  pages={234--241},
  year={2015},
  organization={Springer}
}

@inproceedings{long2015fully,
  title={Fully convolutional networks for semantic segmentation},
  author={Long, Jonathan and Shelhamer, Evan and Darrell, Trevor},
  booktitle={Proceedings of the IEEE conference on computer vision and pattern recognition},
  pages={3431--3440},
  year={2015}
}

@article{chen2017deeplab,
  title={Deeplab: Semantic image segmentation with deep convolutional nets, atrous convolution, and fully connected crfs},
  author={Chen, Liang-Chieh and Papandreou, George and Kokkinos, Iasonas and Murphy, Kevin and Yuille, Alan L},
  journal={IEEE transactions on pattern analysis and machine intelligence},
  volume={40},
  number={4},
  pages={834--848},
  year={2017},
  publisher={IEEE}
}

@article{manfreda2024advancing,
  title={Advancing river monitoring using image-based techniques: challenges and opportunities},
  author={Manfreda, Salvatore and Miglino, Domenico and Saddi, Khim Cathleen and Jomaa, Seifeddine and Eltner, Anette and Perks, Matthew and Pe{\~n}a-Haro, Salvador and Bogaard, Thom and Van Emmerik, Tim HM and Mariani, Stefano and others},
  journal={Hydrological Sciences Journal},
  volume={69},
  number={6},
  pages={657--677},
  year={2024},
  publisher={Taylor \& Francis}
}

@inproceedings{bastani2023satlaspretrain,
  title={Satlaspretrain: A large-scale dataset for remote sensing image understanding},
  author={Bastani, Favyen and Wolters, Piper and Gupta, Ritwik and Ferdinando, Joe and Kembhavi, Aniruddha},
  booktitle={Proceedings of the IEEE/CVF International Conference on Computer Vision},
  pages={16772--16782},
  year={2023}
}

@article{jakubik2310foundation_prithvi,
  title={Foundation models for generalist geospatial artificial intelligence, 2023},
  author={Jakubik, Johannes and Roy, S and Phillips, CE and Fraccaro, P and Godwin, D and Zadrozny, B and Szwarcman, D and Gomes, C and Nyirjesy, G and Edwards, B and others},
  journal={URL https://arxiv. org/abs/2310.18660},
  year={2023},
}

@inproceedings{manas2021seasonal_seco,
  title={Seasonal contrast: Unsupervised pre-training from uncurated remote sensing data},
  author={Manas, Oscar and Lacoste, Alexandre and Gir{\'o}-i-Nieto, Xavier and Vazquez, David and Rodriguez, Pau},
  booktitle={Proceedings of the IEEE/CVF International Conference on Computer Vision},
  pages={9414--9423},
  year={2021}
}

@inproceedings{daroya2025wildsat,
  title={WildSAT: Learning satellite image representations from wildlife observations},
  author={Daroya, Rangel and Cole, Elijah and Mac Aodha, Oisin and Van Horn, Grant and Maji, Subhransu},
  booktitle={Proceedings of the IEEE/CVF International Conference on Computer Vision},
  year={2025}
}

@article{yu2019semantic,
  title={Semantic face hallucination: Super-resolving very low-resolution face images with supplementary attributes},
  author={Yu, Xin and Fernando, Basura and Hartley, Richard and Porikli, Fatih},
  journal={IEEE transactions on pattern analysis and machine intelligence},
  volume={42},
  number={11},
  pages={2926--2943},
  year={2019},
  publisher={IEEE}
}

@inproceedings{yu2017hallucinating,
  title={Hallucinating very low-resolution unaligned and noisy face images by transformative discriminative autoencoders},
  author={Yu, Xin and Porikli, Fatih},
  booktitle={Proceedings of the IEEE conference on computer vision and pattern recognition},
  pages={3760--3768},
  year={2017}
}

@article{martens2019super,
  title={Super-resolution of PROBA-V images using convolutional neural networks},
  author={M{\"a}rtens, Marcus and Izzo, Dario and Krzic, Andrej and Cox, Dani{\"e}l},
  journal={Astrodynamics},
  volume={3},
  number={4},
  pages={387--402},
  year={2019},
  publisher={Springer}
}

@article{cornebise2022open,
  title={Open high-resolution satellite imagery: The worldstrat dataset--with application to super-resolution},
  author={Cornebise, Julien and Or{\v{s}}oli{\'c}, Ivan and Kalaitzis, Freddie},
  journal={Advances in Neural Information Processing Systems},
  volume={35},
  pages={25979--25991},
  year={2022}
}

@article{li2003remote,
  title={Remote sensing of suspended sediments and shallow coastal waters},
  author={Li, Rong-Rong and Gao, Bo-Cai},
  journal={IEEE Transactions on Geoscience and Remote Sensing},
  volume={41},
  number={3},
  pages={559--566},
  year={2003},
  publisher={IEEE}
}

@article{gardner2023human,
  title={Human activities change suspended sediment concentration along rivers},
  author={Gardner, John and Pavelsky, Tamlin and Topp, Simon and Yang, Xiao and Ross, Matthew RV and Cohen, Sagy},
  journal={Environmental Research Letters},
  volume={18},
  number={6},
  pages={064032},
  year={2023},
  publisher={IOP Publishing}
}

@article{dethier2020toward,
  title={Toward improved accuracy of remote sensing approaches for quantifying suspended sediment: Implications for suspended-sediment monitoring},
  author={Dethier, EN and Renshaw, CE and Magilligan, FJ},
  journal={Journal of Geophysical Research: Earth Surface},
  volume={125},
  number={7},
  pages={e2019JF005033},
  year={2020},
  publisher={Wiley Online Library}
}

@article{li2010multi,
  title={A multi-frame image super-resolution method},
  author={Li, Xuelong and Hu, Yanting and Gao, Xinbo and Tao, Dacheng and Ning, Beijia},
  journal={Signal Processing},
  volume={90},
  number={2},
  pages={405--414},
  year={2010},
  publisher={Elsevier}
}

@inproceedings{di2025qmambabsr,
  title={Qmambabsr: Burst image super-resolution with query state space model},
  author={Di, Xin and Peng, Long and Xia, Peizhe and Li, Wenbo and Pei, Renjing and Cao, Yang and Wang, Yang and Zha, Zheng-Jun},
  booktitle={Proceedings of the Computer Vision and Pattern Recognition Conference},
  pages={23080--23090},
  year={2025}
}

@inproceedings{lu2022transformer,
  title={Transformer for single image super-resolution},
  author={Lu, Zhisheng and Li, Juncheng and Liu, Hong and Huang, Chaoyan and Zhang, Linlin and Zeng, Tieyong},
  booktitle={Proceedings of the IEEE/CVF conference on computer vision and pattern recognition},
  pages={457--466},
  year={2022}
}

@inproceedings{dudhane2023burstormer,
  title={Burstormer: Burst image restoration and enhancement transformer},
  author={Dudhane, Akshay and Zamir, Syed Waqas and Khan, Salman and Khan, Fahad Shahbaz and Yang, Ming-Hsuan},
  booktitle={2023 IEEE/CVF Conference on Computer Vision and Pattern Recognition (CVPR)},
  pages={5703--5712},
  year={2023},
  organization={IEEE}
}

@inproceedings{ranftl2021vision,
  title={Vision transformers for dense prediction},
  author={Ranftl, Ren{\'e} and Bochkovskiy, Alexey and Koltun, Vladlen},
  booktitle={Proceedings of the IEEE/CVF international conference on computer vision},
  pages={12179--12188},
  year={2021}
}

@inproceedings{wang2018esrgan,
  title={Esrgan: Enhanced super-resolution generative adversarial networks},
  author={Wang, Xintao and Yu, Ke and Wu, Shixiang and Gu, Jinjin and Liu, Yihao and Dong, Chao and Qiao, Yu and Change Loy, Chen},
  booktitle={Proceedings of the European conference on computer vision (ECCV) workshops},
  pages={0--0},
  year={2018}
}

@inproceedings{heDeepResidualLearning2016,
  title={Deep residual learning for image recognition},
  author={He, Kaiming and Zhang, Xiangyu and Ren, Shaoqing and Sun, Jian},
  booktitle={Proceedings of the IEEE conference on computer vision and pattern recognition},
  pages={770--778},
  year={2016}
}

@inproceedings{sandler2018mobilenetv2,
  title={Mobilenetv2: Inverted residuals and linear bottlenecks},
  author={Sandler, Mark and Howard, Andrew and Zhu, Menglong and Zhmoginov, Andrey and Chen, Liang-Chieh},
  booktitle={Proceedings of the IEEE conference on computer vision and pattern recognition},
  pages={4510--4520},
  year={2018}
}

@article{dosovitskiy2020vit,
  title={An Image is Worth 16x16 Words: Transformers for Image Recognition at Scale},
  author={Dosovitskiy, Alexey and Beyer, Lucas and Kolesnikov, Alexander and Weissenborn, Dirk and Zhai, Xiaohua and Unterthiner, Thomas and  Dehghani, Mostafa and Minderer, Matthias and Heigold, Georg and Gelly, Sylvain and Uszkoreit, Jakob and Houlsby, Neil},
  journal={International Conference on Learning Representations (ICLR)},
  year={2021}
}

@inproceedings{liu2021swin,
  title={Swin transformer: Hierarchical vision transformer using shifted windows},
  author={Liu, Ze and Lin, Yutong and Cao, Yue and Hu, Han and Wei, Yixuan and Zhang, Zheng and Lin, Stephen and Guo, Baining},
  booktitle={Proceedings of the IEEE/CVF international conference on computer vision},
  year={2021}
}

@inproceedings{deng2009imagenet,
  title={Imagenet: A large-scale hierarchical image database},
  author={Deng, Jia and Dong, Wei and Socher, Richard and Li, Li-Jia and Li, Kai and Fei-Fei, Li},
  booktitle={2009 IEEE conference on computer vision and pattern recognition},
  pages={248--255},
  year={2009},
  organization={Ieee}
}

@inproceedings{chen2021empirical_mocov3,
  title={An empirical study of training self-supervised vision transformers},
  author={Chen, Xinlei and Xie, Saining and He, Kaiming},
  booktitle={Proceedings of the IEEE/CVF international conference on computer vision},
  pages={9640--9649},
  year={2021}
}

@inproceedings{caron2021emerging,
  title={Emerging properties in self-supervised vision transformers},
  author={Caron, Mathilde and Touvron, Hugo and Misra, Ishan and J{\'e}gou, Herv{\'e} and Mairal, Julien and Bojanowski, Piotr and Joulin, Armand},
  booktitle={Proceedings of the IEEE/CVF international conference on computer vision},
  pages={9650--9660},
  year={2021}
}

@inproceedings{radford2021learning_clip,
  title={Learning transferable visual models from natural language supervision},
  author={Radford, Alec and Kim, Jong Wook and Hallacy, Chris and Ramesh, Aditya and Goh, Gabriel and Agarwal, Sandhini and Sastry, Girish and Askell, Amanda and Mishkin, Pamela and Clark, Jack and others},
  booktitle={International conference on machine learning},
  pages={8748--8763},
  year={2021},
  organization={PMLR}
}

@article{zanaga2022esa,
  title={ESA WorldCover 10 m 2021 v200},
  author={Zanaga, Daniele and Van De Kerchove, Ruben and Daems, Dirk and De Keersmaecker, Wanda and Brockmann, Carsten and Kirches, Grit and Wevers, Jan and Cartus, Oliver and Santoro, Maurizio and Fritz, Steffen and others},
  year={2022},
  publisher={Zenodo}
}

@article{qiu2019fmask,
  title={Fmask 4.0: Improved cloud and cloud shadow detection in Landsats 4--8 and Sentinel-2 imagery},
  author={Qiu, Shi and Zhu, Zhe and He, Binbin},
  journal={Remote Sensing of Environment},
  volume={231},
  pages={111205},
  year={2019},
  publisher={Elsevier}
}

@article{kruse2015validation,
  title={Validation of DigitalGlobe WorldView-3 Earth imaging satellite shortwave infrared bands for mineral mapping},
  author={Kruse, Fred A and Baugh, William M and Perry, Sandra L},
  journal={Journal of Applied Remote Sensing},
  volume={9},
  number={1},
  pages={096044--096044},
  year={2015},
  publisher={Society of Photo-Optical Instrumentation Engineers}
}

@inproceedings{crespi2010geoeye,
  title={GeoEye-1: Analysis of radiometric and geometric capability},
  author={Crespi, Mattia and Colosimo, Gabriele and De Vendictis, Laura and Fratarcangeli, Francesca and Pieralice, Francesca},
  booktitle={International Conference on Personal Satellite Services},
  pages={354--369},
  year={2010},
  organization={Springer}
}

@article{langhorst2024global,
  title={Global cloud biases in optical satellite remote sensing of rivers},
  author={Langhorst, Theodore and Andreadis, Konstantinos M and Allen, George H},
  journal={Geophysical Research Letters},
  volume={51},
  number={16},
  pages={e2024GL110085},
  year={2024},
  publisher={Wiley Online Library}
}

@article{kuriqi2021ecological,
  title={Ecological impacts of run-of-river hydropower plants—Current status and future prospects on the brink of energy transition},
  author={Kuriqi, Alban and Pinheiro, Ant{\'o}nio N and Sordo-Ward, Alvaro and Bejarano, Mar{\'\i}a D and Garrote, Luis},
  journal={Renewable and Sustainable Energy Reviews},
  volume={142},
  pages={110833},
  year={2021},
  publisher={Elsevier}
}

@article{anderson2015impacts,
  title={The impacts of ‘run-of-river’hydropower on the physical and ecological condition of rivers},
  author={Anderson, David and Moggridge, Helen and Warren, Philip and Shucksmith, James},
  journal={Water and Environment Journal},
  volume={29},
  number={2},
  pages={268--276},
  year={2015},
  publisher={Wiley Online Library}
}

@article{li2022integrating,
  title={Integrating river health into the supply and demand management framework for river basin ecosystem services},
  author={Li, Tianjiao and Wang, Huimin and Fang, Zhou and Liu, Gang and Zhang, Fan and Zhang, Haitao and Li, Xuxia},
  journal={Sustainable Production and Consumption},
  volume={33},
  pages={189--202},
  year={2022},
  publisher={Elsevier}
}

@article{wurbs2005modeling,
  title={Modeling river/reservoir system management, water allocation, and supply reliability},
  author={Wurbs, Ralph A},
  journal={Journal of Hydrology},
  volume={300},
  number={1-4},
  pages={100--113},
  year={2005},
  publisher={Elsevier}
}

@article{kim2024evaluation,
  title={Evaluation of Deep Learning-Based Water Bodies and Flooded Area Detection with Nanosatellites: The PlanetScope Satellite Imageries and HRNet Model},
  author={Kim, Wanyub and Cho, Shinhyeon and Jeong, Junhyuk and Kim, Yeji and Kim, Hyun Ok and Choi, Minha},
  journal={Korean Journal of Remote Sensing},
  volume={40},
  number={5},
  pages={617--627},
  year={2024},
  publisher={Korean Society of Remote Sensing}
}

@article{bjerklie2018satellite,
  title={Satellite remote sensing estimation of river discharge: Application to the Yukon River Alaska},
  author={Bjerklie, David M and Birkett, Charon M and Jones, John W and Carabajal, Claudia and Rover, Jennifer A and Fulton, John W and Garambois, Pierre-Andr{\'e}},
  journal={Journal of hydrology},
  volume={561},
  pages={1000--1018},
  year={2018},
  publisher={Elsevier}
}

@article{hagemann2017bam,
  title={BAM: Bayesian AMHG-Manning inference of discharge using remotely sensed stream width, slope, and height},
  author={Hagemann, MW and Gleason, CJ and Durand, MT},
  journal={Water Resources Research},
  volume={53},
  number={11},
  pages={9692--9707},
  year={2017},
  publisher={Wiley Online Library}
}

@article{ad2001global,
  title={Global water data: A newly endangered species},
  author={Ad Hoc Group and V{\"o}r{\"o}smarty, C and Askew, A and Grabs, W and Barry, RG and Birkett, C and D{\"o}ll, P and Goodison, B and Hall, A and Jenne, R and others},
  journal={Eos, Transactions American Geophysical Union},
  volume={82},
  number={5},
  pages={54--58},
  year={2001},
  publisher={Wiley Online Library}
}

@article{gleason2017crossing,
  title={Crossing the (watershed) divide: Satellite data and the changing politics of international river basins},
  author={Gleason, Colin J and Hamdan, Ali N},
  journal={The Geographical Journal},
  volume={183},
  number={1},
  pages={2--15},
  year={2017},
  publisher={Wiley Online Library}
}

@article{chen2005water,
  title={Water demand management: A case study of the Heihe River Basin in China},
  author={Chen, Yan and Zhang, Dunqiang and Sun, Yangbo and Liu, Xinai and Wang, Nianzhong and Savenije, Hubert HG},
  journal={Physics and Chemistry of the Earth, Parts A/B/C},
  volume={30},
  number={6-7},
  pages={408--419},
  year={2005},
  publisher={Elsevier}
}

@article{patil2018understanding,
  title={Understanding hydro-ecological surprises for riverine ecosystem management},
  author={Patil, Rupesh and Wei, Yongping and Pullar, David and Shulmeister, James},
  journal={Current Opinion in Environmental Sustainability},
  volume={33},
  pages={142--150},
  year={2018},
  publisher={Elsevier}
}

@article{yang2019rivwidthcloud,
  title={RivWidthCloud: An automated Google Earth Engine algorithm for river width extraction from remotely sensed imagery},
  author={Yang, Xiao and Pavelsky, Tamlin M and Allen, George H and Donchyts, Gennadii},
  journal={IEEE Geoscience and Remote Sensing Letters},
  volume={17},
  number={2},
  pages={217--221},
  year={2019},
  publisher={IEEE}
}
}

\appendix 

\maketitlesupplementary
\setcounter{page}{1}

\setcounter{table}{0}
\renewcommand{\thetable}{A\arabic{table}}
\setcounter{figure}{0}
\renewcommand{\thefigure}{A\arabic{figure}}

\section{Additional Results}

\subsection{Segmentation performance.}

\begin{figure*}[!h]
    \centering
    \includegraphics[width=\linewidth]{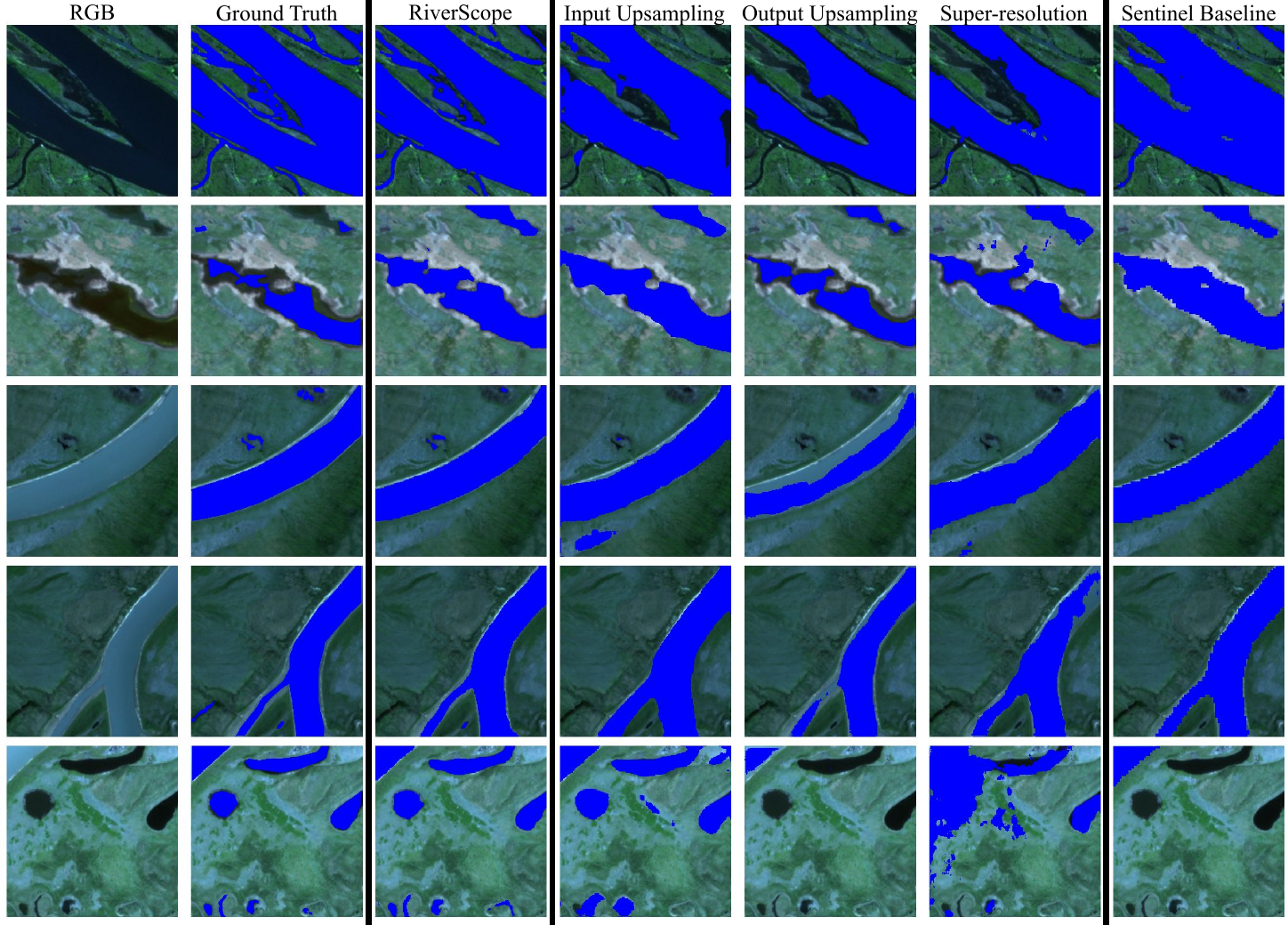}
    \vspace{-18pt}
    \caption{\textbf{Visualization of segmentation outputs from different methods}.}
    \label{fig:supp-segmentation-viz-wbaselines}
\end{figure*}

\paragraph{Segmentation outputs compared to baselines.}
\cref{fig:supp-segmentation-viz-wbaselines} shows visualization of segmentation outputs from different methods.
The HR model (RiverScope) outputs segmentations with precise boundaries and are very close to the ground truth labels.
In contrast, the sentinel baseline outputs ragged segmentations with less defined boundaries.
Input upsampling is generally more consistent with the ground truth than output upsampling and super-resolution.
Output upsampling tends to miss fine-grained details, which is likely due to the upsampling only being done after the model has already made its predictions---the model is operating on LR images, which naturally loses details.
Super-resolution, on the other hand, can hallucinate details (last row of \cref{fig:supp-segmentation-viz-wbaselines}) causing it to make mistakes.

\begin{figure*}[!h]
    \centering
    \includegraphics[width=\linewidth]{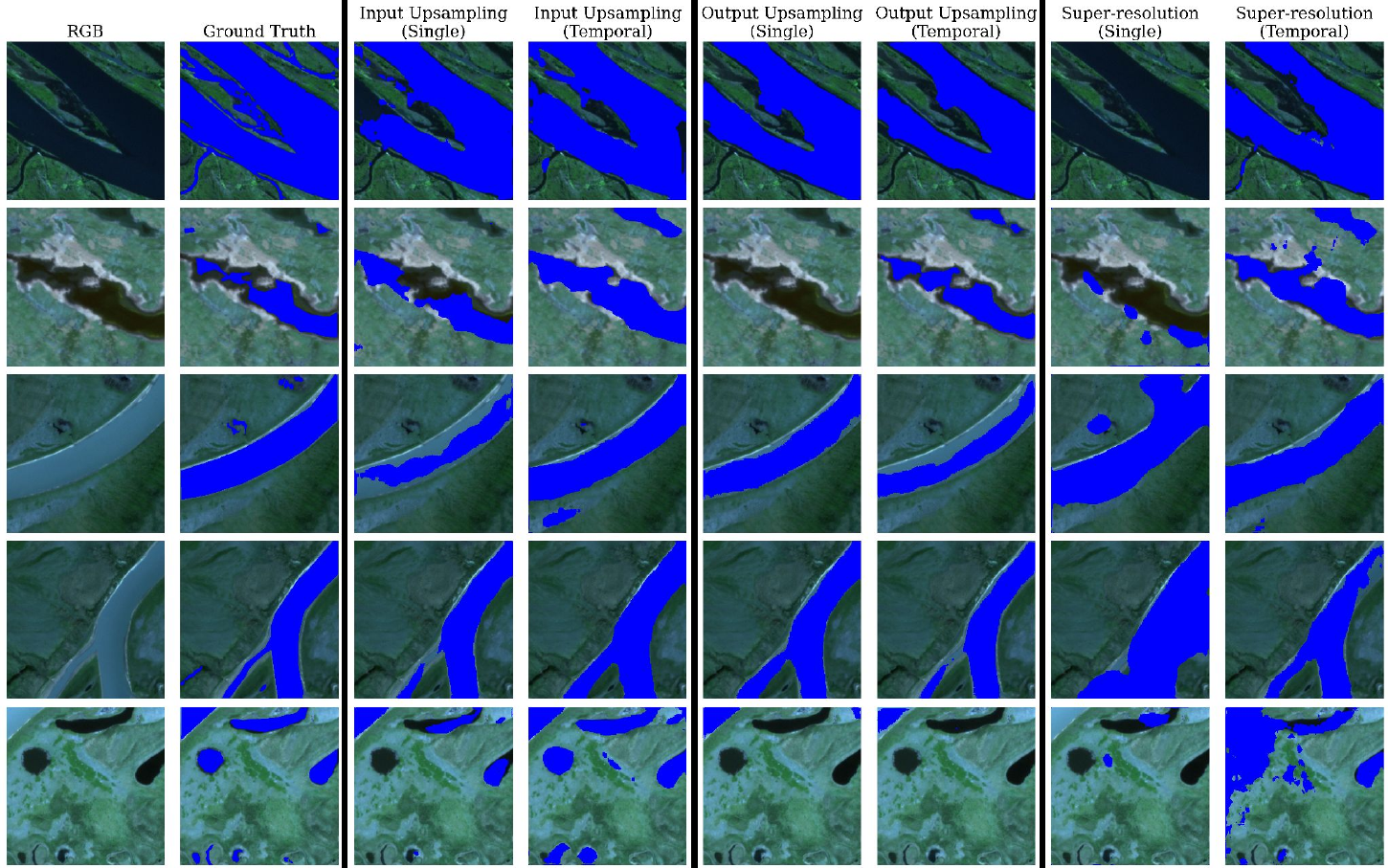}
    \vspace{-18pt}
    \caption{\textbf{Visualization of segmentation outputs when using single versus temporal inputs}.}
    \label{fig:supp-segmentation-single-vs-temporal}
\end{figure*}

\paragraph{Segmentation comparison of single and temporal inputs.}
\cref{fig:supp-segmentation-single-vs-temporal} shows the qualitative comparison of different methods that use single or temporal inputs.
In general, more details can be distinguished when using temporal inputs, resulting in better segmentation quality.
Using temporal inputs helps the models be less susceptible to atmospheric noise like clouds.

\begin{figure*}[!h]
    \centering
    \includegraphics[width=\linewidth]{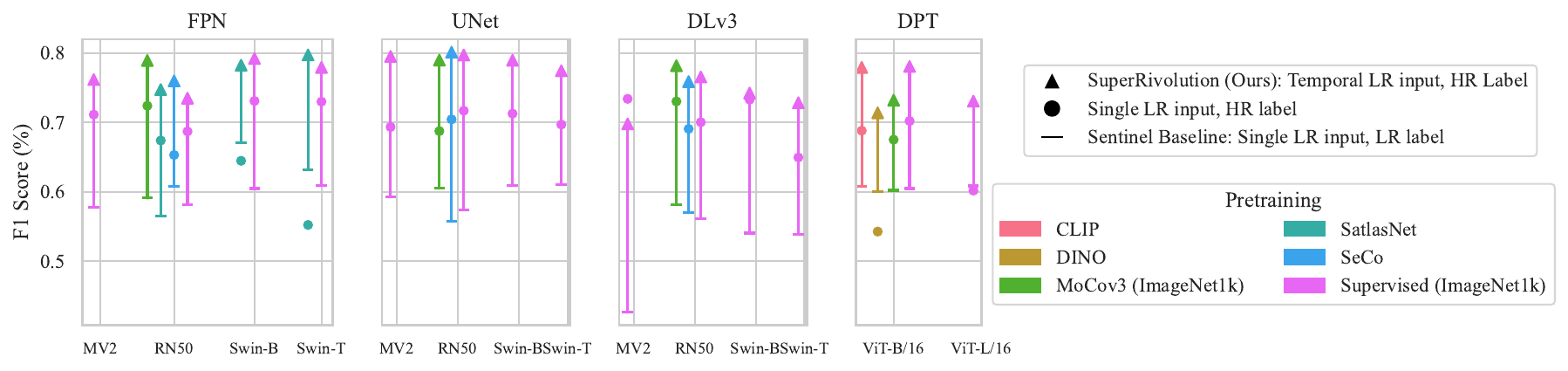}
    \vspace{-18pt}
    \caption{\textbf{Input upsampling segmentation metrics across different pretraining methods}.}
    \label{fig:supp-segmentation-input-upsampling}
\end{figure*}

\begin{figure*}[!h]
    \centering
    \includegraphics[width=\linewidth]{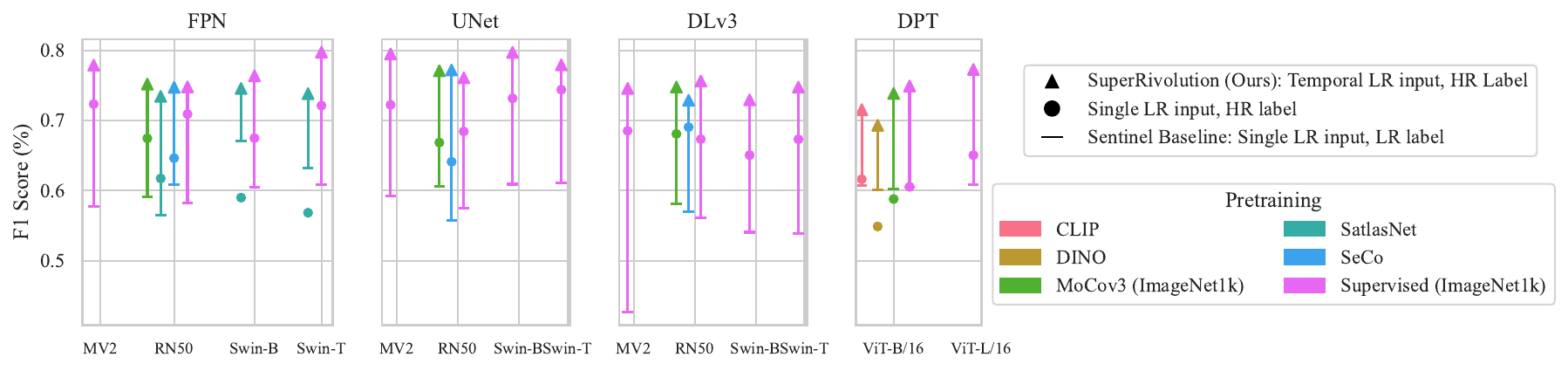}
    \vspace{-18pt}
    \caption{\textbf{Output upsampling segmentation metrics across different pretraining methods}.}
    \label{fig:supp-segmentation-output-upsampling}
\end{figure*}

\begin{figure*}[!h]
    \centering
    \includegraphics[width=\linewidth]{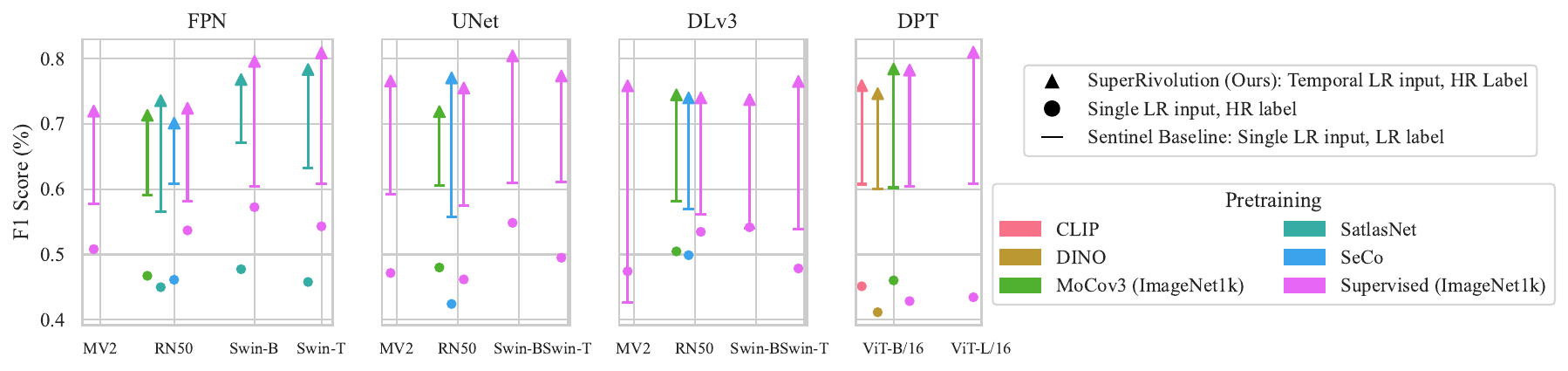}
    \vspace{-18pt}
    \caption{\textbf{Super-resolution segmentation metrics across different pretraining methods}.}
    \label{fig:supp-segmentation-sr}
\end{figure*}

\paragraph{Results across pretraining methods.}
\cref{fig:supp-segmentation-input-upsampling}, \cref{fig:supp-segmentation-output-upsampling}, and \cref{fig:supp-segmentation-sr} show segmentation performance across different pretraining methods, segmentation models, and backbones. Similar to trends observed in \cref{fig:segmentation-metrics}, using temporal inputs with SuperRivolution result in better F1 score than the low resolution Sentinel baseline (horizontal markers).

\subsection{Cloud cover impact on performance}
\begin{figure*}[!h]
    \centering
    \includegraphics[width=\linewidth]{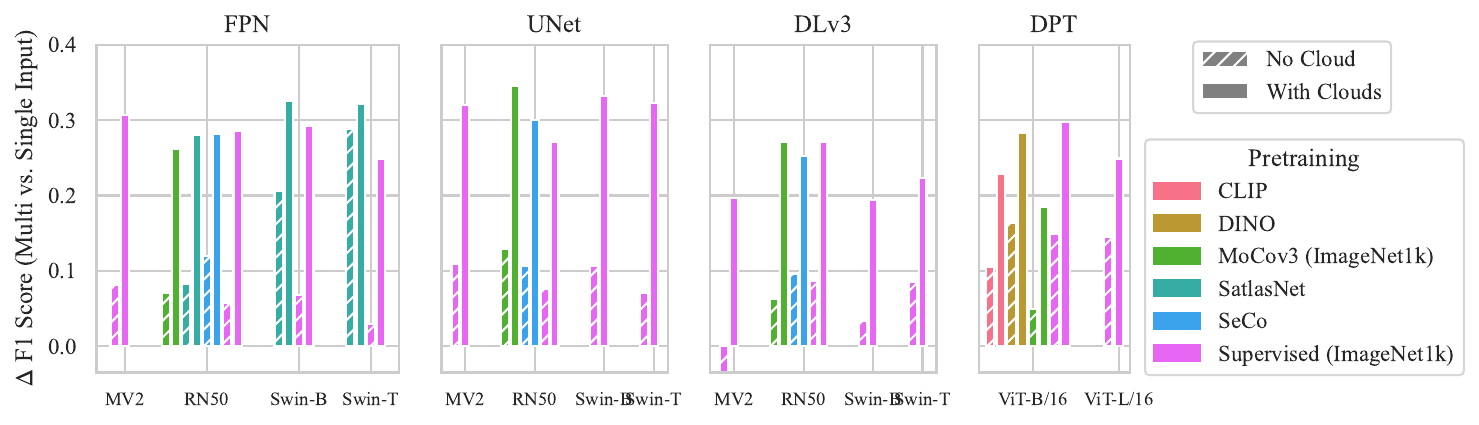}
    \vspace{-18pt}
    \caption{\textbf{Impact of cloud cover on temporal input performance gain for input upsampling}.}
    \label{fig:supp-cloudy-metrics-input-upsampling}
\end{figure*}

\begin{figure*}[!h]
    \centering
    \includegraphics[width=\linewidth]{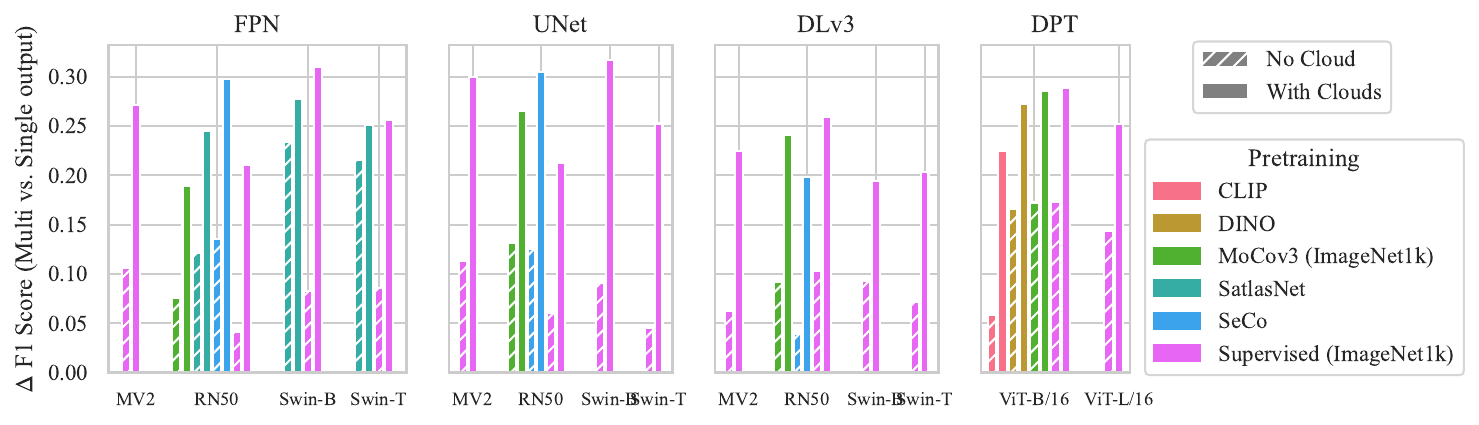}
    \vspace{-18pt}
    \caption{\textbf{Impact of cloud cover on temporal input performance gain for output upsampling}.}
    \label{fig:supp-cloudy-metrics-output-upsampling}
\end{figure*}

\begin{figure*}[!h]
    \centering
    \includegraphics[width=\linewidth]{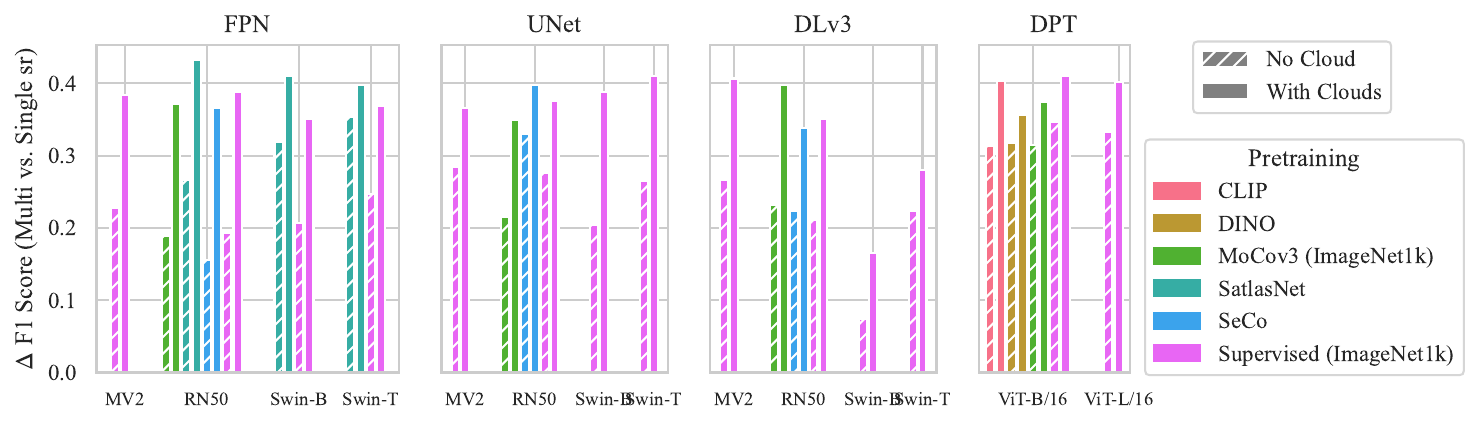}
    \vspace{-18pt}
    \caption{\textbf{Impact of cloud cover on temporal input performance gain for super-resolution}.}
    \label{fig:supp-cloudy-metrics-sr}
    \vspace{-9pt}
\end{figure*}

\paragraph{Results across pretraining methods.}
\cref{fig:supp-cloudy-metrics-input-upsampling}, \cref{fig:supp-cloudy-metrics-output-upsampling}, and \cref{fig:supp-cloudy-metrics-sr} show the impact of performance when using a temporal input versus a single image input.
The three plots show performance across different backbones, segmentation models, and pretraining methods.
Similar to results in \cref{fig:segmentation-cloud-effect}, we generally see high performance gain on cloudy images.

\begin{figure*}[!h]
    \centering
    \includegraphics[width=0.9\linewidth]{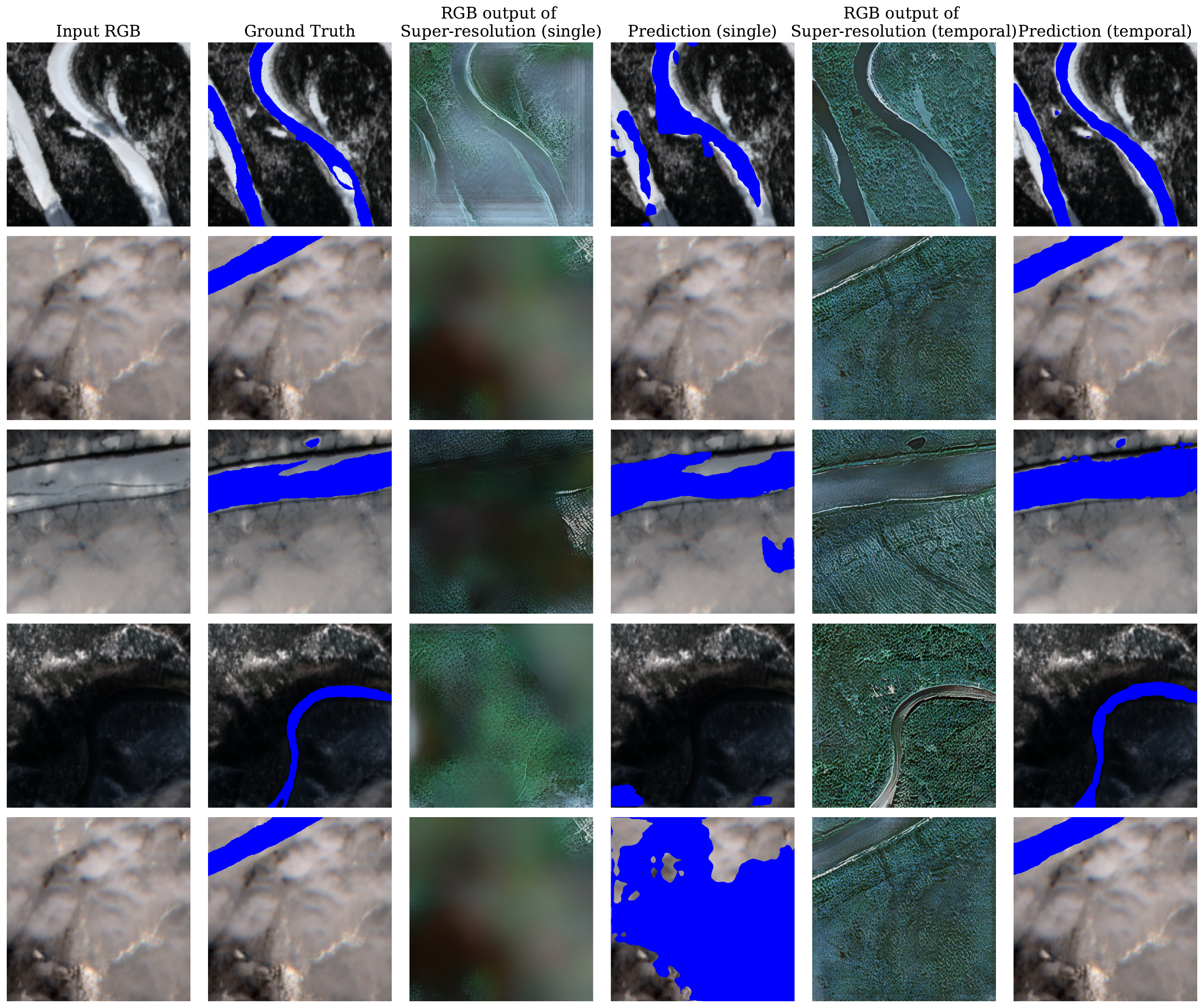}
    \vspace{-12pt}
    \caption{\textbf{Visualization of the impact of noisy and cloudy images when applying the super-resolution model}.}
    \label{fig:supp-cloudy-samples-superres}
    \vspace{-12pt}
\end{figure*}

\paragraph{Cloud cover impact on super-resolution.}
\cref{fig:supp-cloudy-samples-superres} shows that using temporal input for an otherwise noisy or cloudy single image significantly improves the performance.
In particular, when applying the super-resolution model ESRGAN~\cite{wang2018esrgan} the performance significantly improves for temporal inputs over single image inputs.
By ensembling the model output across multiple inputs, the effect of the noise is minimized.

\subsection{River width estimation errors.}
\begin{figure}[!t]
    \centering
    \includegraphics[width=0.8\linewidth]{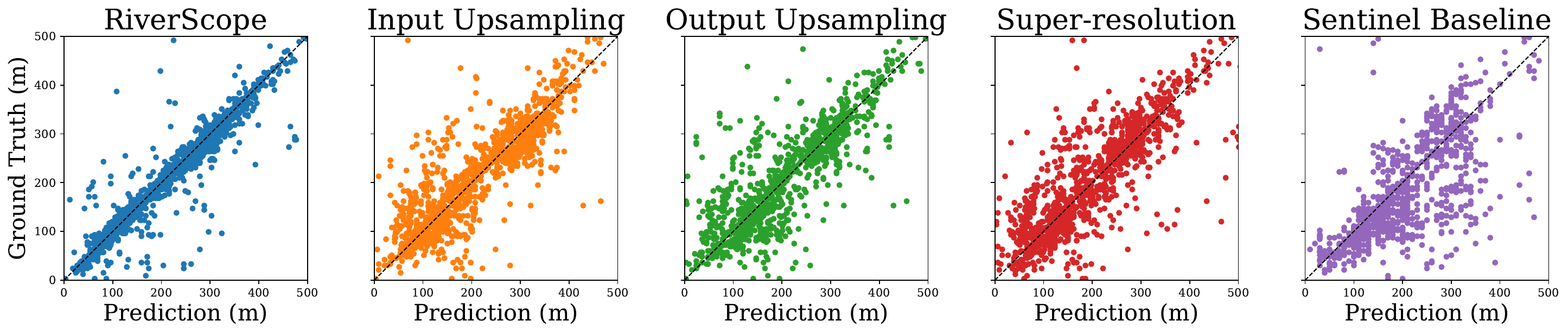}
    \vspace{-9pt}
    \caption{\textbf{Distribution of predicted widths from different methods}. Temporal inputs are used for input upsampling, output upsampling, and super-resolution. Single images are used for RiverScope and Sentinel Baseline. Sentinel baseline predictions appear to be more skewed than the SuperRivolution methods.
    }
    \label{fig:supp-river-width-scatter}
    \vspace{-9pt}
\end{figure}

\paragraph{Distribution of width estimates.}
\cref{fig:supp-river-width-scatter} shows the distribution of river width estimates of the different methods. Models that take multiple images as input are generally less biased with the prediction clustering closely to the $y=x$ line.

\begin{figure*}[!h]
    \centering
    \includegraphics[width=\linewidth]{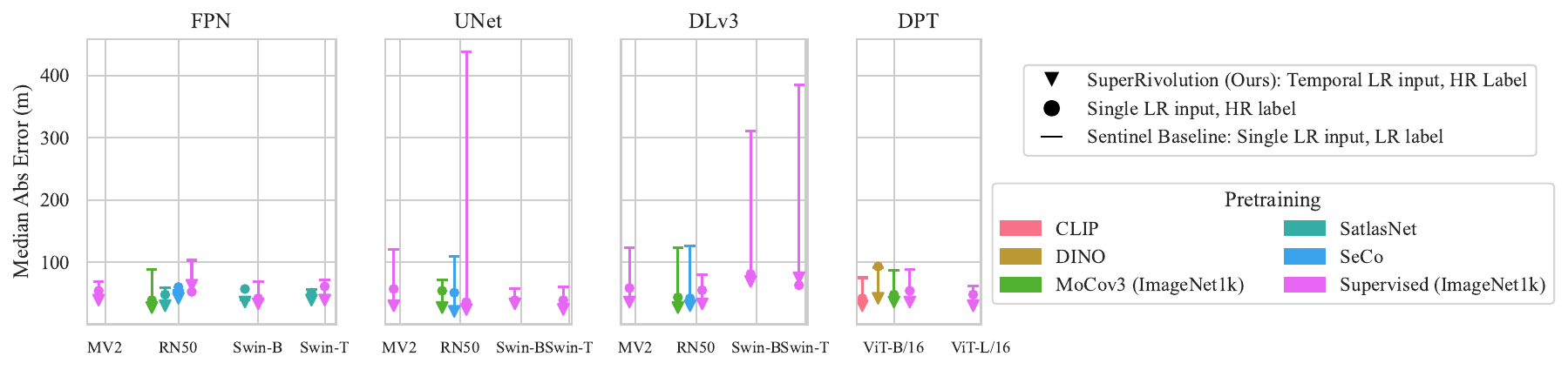}
    \caption{\textbf{Input upsampling river width estimation errors across different pretraining methods}.}
    \label{fig:supp-river-width-input-upsampling}
\end{figure*}

\begin{figure*}[!h]
    \centering
    \includegraphics[width=\linewidth]{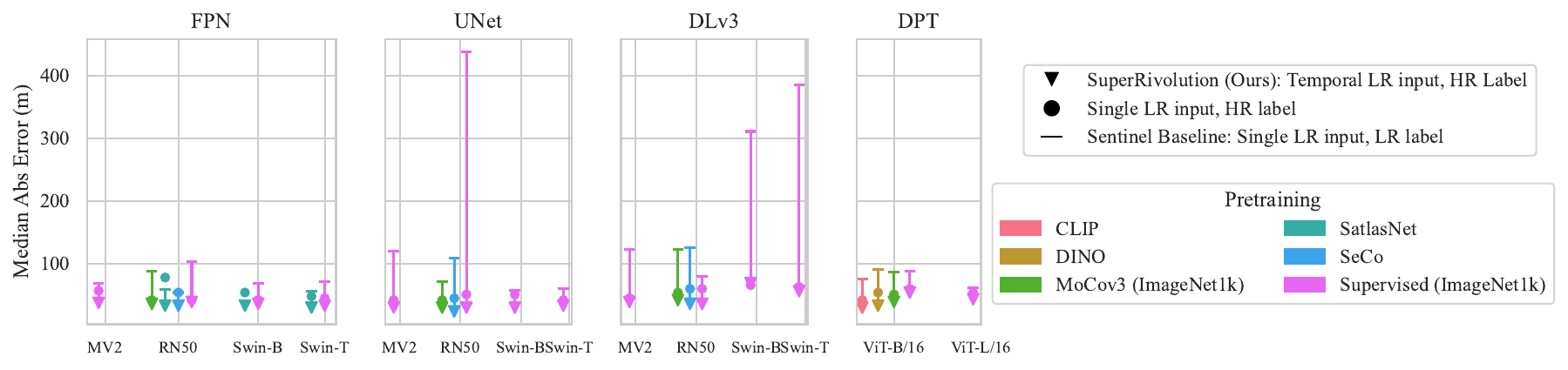}
    \caption{\textbf{Output upsampling river width estimation errors across different pretraining methods}.}
    \label{fig:supp-river-width-output-upsampling}
\end{figure*}

\begin{figure*}[!h]
    \centering
    \includegraphics[width=\linewidth]{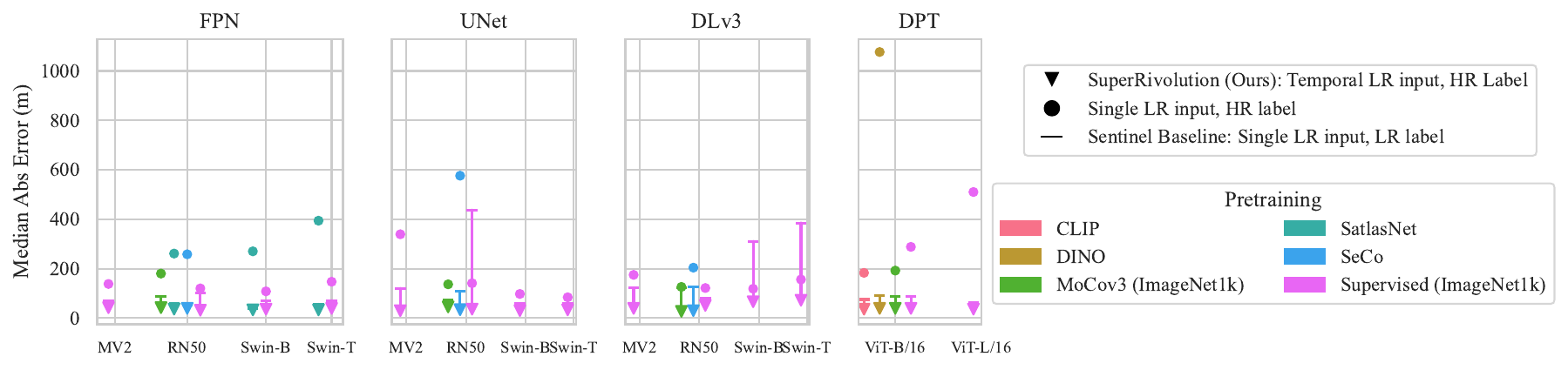}
    \caption{\textbf{Super-resolution river width estimation errors across different pretraining methods}.}
    \label{fig:supp-river-width-sr}
\end{figure*}

\paragraph{Results across pretraining methods.}
\cref{fig:supp-river-width-input-upsampling}, \cref{fig:supp-river-width-output-upsampling}, and \cref{fig:supp-river-width-sr} show the width estimation performance across different backbones, segmentation models, and pretraining methods. Consistent with the results in \cref{fig:river-width-metrics}, the performance improves when temporal inputs are used as input.

\end{document}